\documentclass[useAMS,usenatbib]{mn2e}
\usepackage{amsmath,natbib,lscape,graphicx,natbib,amssymb,times,multirow,epsfig}

\title[Orbital motion monitoring of A-type star multiples]{The VAST
  Survey -- II. Orbital motion monitoring of A-type star multiples}

\author[De Rosa et al.]
{R. J. De Rosa$^1$\thanks{E-mail: derosa@astro.ex.ac.uk}\thanks{Based
    on observations obtained at the Canada-France-Hawaii Telescope
    (CFHT) under programme IDs 2001AF11,
    2008AC22, 2009BC06, 2010AC14, 2011AC11.
    Based on observations made with ESO
    Telescopes at the La Silla Paranal Observatory under programme IDs
    272.D-5068(A), 074.D-0180(A),
    076.C-0270(A), 076.D-0108(A),
    077.D-0147(A), 080.D-0348(A).
    Based on observations
    obtained at the Gemini Observatory under programme IDs
    GN2008A-Q-74, GN-2008B-Q-119, GN-2010A-Q-75.}, J. Patience$^1$,
  A. Vigan$^1$, P. A. Wilson$^1$, A. Schneider$^2$,  N. J. McConnell$^3$,\newauthor
  S. J. Wiktorowicz$^{3,4}$, C. Marois$^5$, I. Song$^2$,  B. Macintosh$^6$,
  J. R. Graham$^{3,7}$, M. S. Bessell$^8$,\newauthor R. Doyon$^9$, \&
  O. Lai$^{10}$\\
$^1$ College of Engineering, Mathematics and Physical Sciences,
Physics Building, University of Exeter, Stocker Road, Exeter, EX4 4QL,
UK\\
$^2$ Physics and Astronomy, University of Georgia, 240 Physics, Athens, GA 30602,
USA\\
$^3$ Department of Astronomy, University of California at Berkeley,
Berkeley, CA 94720, USA\\
$^4$ Department of Astronomy, University of California at Santa Cruz,
Santa Cruz, CA 95064, USA\\
$^5$ NRC Herzberg Institute of Astrophysics, 5071 West Saanich Road, Victoria,
BC, V9E 2E7, Canada\\
$^6$ Institute of Geophysics and Planetary Physics, Lawrence Livermore National Laboratory, 7000 East Ave, Livermore, CA 94550,
USA\\
$^7$ Dunlap Institute for Astronomy and Astrophysics, University of Toronto, 50 St. George Street, Toronto, ON, M55 3H8, Canada\\
$^8$ Mount Stromlo and Siding Spring Observatories, Institute of Advanced
Studies, The Australian National University, Weston Creek P.O., ACT 2611,
Australia\\
$^9$ D\'{e}pt de Physique, Universit\'{e} de Montr\'{e}al, C.P. 6128, Succ. Centre-Ville, Montr\'{e}al, QC,
H3C 3J7, Canada\\
$^{10}$ Canda-France-Hawaii Telescope, 65-1238 Mamalahoa Highway,
Kamuela, HI 96745, USA}
\date{Accepted 2011 December 15. Received 2011 December 13; in original
  form 2011 November 29}
\pagerange{\pageref{firstpage}--\pageref{lastpage}} \pubyear{2011}
\begin{document}
\label{firstpage}

\maketitle

\begin{abstract}
As a part of our ongoing Volume-limited A-Star (VAST) adaptive optics survey, we have
obtained observations of 26 binary systems with projected separations
$<$100 AU, 13 of which have sufficient historical measurements to
allow for refinement of their orbital elements. For each system with
an estimated orbit, the dynamical system mass obtained was compared
with the system mass estimated from mass-magnitude
relations. Discrepancies between the dynamical and theoretical system
mass can be explained by the presence of a previously unresolved
spectroscopic component, or by a non-solar metallicity of the
system. Using this approach to infer the presence of additional
companions, a lower limit to the fraction of binaries, triples, and
quadruples can be estimated as 39, 46, and 15 per cent, for systems
with at least one companion within 100 AU. The fraction of multiple
systems with three or more components shows a relative increase
compared to the fraction for Solar-type primaries resolved in previous
volume-limited surveys. The observations have also revealed a pair of
potentially young ($<$100 Myr) M-dwarf companions, which would make an
ideal benchmark for the theoretical models during the pre-Main
Sequence contraction phase for M-dwarfs. In addition to those systems
with orbit fits, we report 13 systems for which further orbital
monitoring observations are required, 11 of which are newly resolved
as a part of the VAST survey.
\end{abstract}

\section{Introduction}
The orbits of binary stars offer one of the few techniques to
determine stellar masses, or masses and radii in cases with a
favorable geometry (e.g. \citealp{Andersen:1991wd}). Orbits of binaries with young
ages are particularly important to test theoretical evolutionary
models, and examples include a double-lined eclipsing binary in Orion
(e.g.\citealp{Stassun:2008fp}). Low mass stars and brown dwarfs also
represent a regime requiring empirical calibration, and the visual
orbits of nearby M-, L-, and T- dwarfs have been used to measure system
masses and compare with theoretical mass-luminosity relations \citep{Dupuy:2009jba}.

Visual orbits also provide a method to search for evidence of
additional components and determine higher order multiplicity, by
identifying systems with dynamical masses significantly in excess of
the theoretical predictions. These visual binaries with an indication
of unresolved companions can be monitored with spectroscopy or
interferometry to determine the properties such as period and mass
ratio of the closer pair and augment the statistics compiled from
catalogues \citep{Tokovinin:2008bl}. The properties of  higher order multiple
systems represent tests of formation scenarios including fragmentation
of cores (e.g. \citealp{Pringle:1989vl,Bonnell:2001tb}) and disks
(e.g. \citealp{Stamatellos:2007be,Kratter:2010gf}) and may be
influenced by processes such as accretion (e.g. \citealp{Bate:2000ir})
and dynamical interactions
(e.g. \citealp{McDonald:1995uw,Lodato:2007ja}) 

In this paper, a subset of the systems resolved by our ongoing
volume-limited A-star (VAST) survey are used to determine
dynamical system masses from orbit fits and to compare the results
with theoretical models and search for additional unresolved stellar
companions. The sample of AO-imaged binaries considered in this study
is detailed in Section 2, and a short summary of the new observations
is given in Section 3. The data analysis, including the AO image
processing to determine the relative positions and the subsequent
orbit determination from the compilation of position data, is
explained in Section 4. Section 5 reports the astrometric results from
the new measurements and the orbital elements and masses based on the
orbit fits. The discussion in Section 6 covers a comparison with
theoretical mass-magnitude relations, an assessment of the higher
order multiplicity, and the identification of a set of targets for
continued monitoring. Finally, Section 7 provides a summary and
future directions for the project.

\section{Sample}
\begin{table*}
\caption{Basic properties of sample members}
{\small \begin{tabular}{ccccccr@{$\pm$}lr@{$\pm$}lr@{$\pm$}l}
HIP&Name&HR&HD&ADS&Spectral&\multicolumn{2}{c}{Distance}&\multicolumn{2}{c}{$V_{\rm T}$}&\multicolumn{2}{c}{$K_{\rm s}$}\\
&&&&&Type&\multicolumn{2}{c}{(pc)}&\multicolumn{2}{c}{(mag)}&\multicolumn{2}{c}{(mag)}\\
\hline
\hline
\multicolumn{12}{c}{Orbit Subsample}\\
\hline
5300&$\upsilon$ Phe&331&6767&&A3IV&57.0&2.0&5.230&0.001&4.78&0.02\\
9480&48 Cas&575&12111&1598&A3IV&35.3&0.6&4.533&0.002&4.08&0.13$^{\rm a}$\\
11569&$\iota$ Cas&707&15089&1860&A5p&40.7&1.3&4.496&0.003&4.25&0.03\\
17954&&1188&23985&2799&A2V+...&56.5&1.8&5.259&0.004&4.81&0.02\\
28614&$\mu$ Ori&2124&40932&4617&&47.5&1.5&4.150&0.002&3.64&0.26\\
36850&Castor&2891/2890&60178J&6175&A2Vm&15.6&0.9&1.590&0.020&1.47&0.03$^{\rm a}$\\
44127&$\iota$ UMa&3569&76644&&A7V&14.51&0.03&3.159&0.002&2.67&0.03$^{\rm b}$\\
47479&&3863&84121&&A3IV&72.6&2.2&5.334&0.003&4.80&0.02\\
76952&$\gamma$ CrB&5849&140436&9757&B9IV+...&44.8&1.0&3.819&0.002&3.67&0.23\\
77660&b Ser&5895&141851&&A3Vn&49.8&0.8&5.112&0.002&4.70&0.02\\
80628&$\upsilon$ Oph&6129&148367&&A3m&41.0&1.5&4.657&0.003&4.17&0.04\\
82321&52 Her&6254&152107&10227&A2Vspe...&55.3&1.0&4.833&0.003&4.57&0.02\\
93506&$\zeta$ Sgr&7194&176687&11950&A2.5Va&27.0&0.6&2.617&0.003&2.29&0.23\\
\hline
\multicolumn{12}{c}{Monitoring Subsample}\\
\hline
128&&&224890&&Am...&70.8&1.7&6.521&0.004&6.02&0.02\\
2381&&118&2696&&A3V&53.1&0.8&5.188&0.003&4.83&0.02\\
2852&BG Cet&151&3326&&A5m...&48.9&0.8&6.093&0.004&5.42&0.01\\
5310&79 Psc&328&6695&&A3V&47.3&1.7&5.581&0.003&5.22&0.02\\
18217&&1192&24141&&A5m&50.5&1.1&5.806&0.004&5.37&0.02\\
29852&&2265&43940&&A2V&61.9&1.0&5.895&0.003&5.44&0.02\\
51384&&4062&89571&&F0IV&40.6&0.6&5.549&0.003&4.85&0.02\\
65241&64 Vir&5040&116235&&A2m&65.9&1.4&5.897&0.003&5.62&0.02\\
66223&&5108&118156&8956&F0IV&69.8&2.0&6.394&0.004&5.88&0.02\\
103298&16 Del&8012&199254&14429&A4V&60.5&1.1&5.555&0.003&5.19&0.02\\
109667&&8464&210739&&A3V&63.5&2.1&6.209&0.005&5.74&0.02\\
110787&$\rho^1$ Cep&8578&213403&&A2m&63.2&0.9&5.857&0.004&5.54&0.03\\
116611&75 Peg&8963&222133&&A1Vn&71.4&1.4&5.483&0.003&5.42&0.02\\
\hline
\multicolumn{8}{l}{a - $K$-band photometry from
  \citet{Ducati:2002tr}}\\
\multicolumn{8}{l}{b - $K$-band photometry from \citet{Morel:1978vm}}
\label{tab1}
\end{tabular} }
\end{table*}
\begin{figure}
\resizebox{0.9\hsize}{!}{{\includegraphics{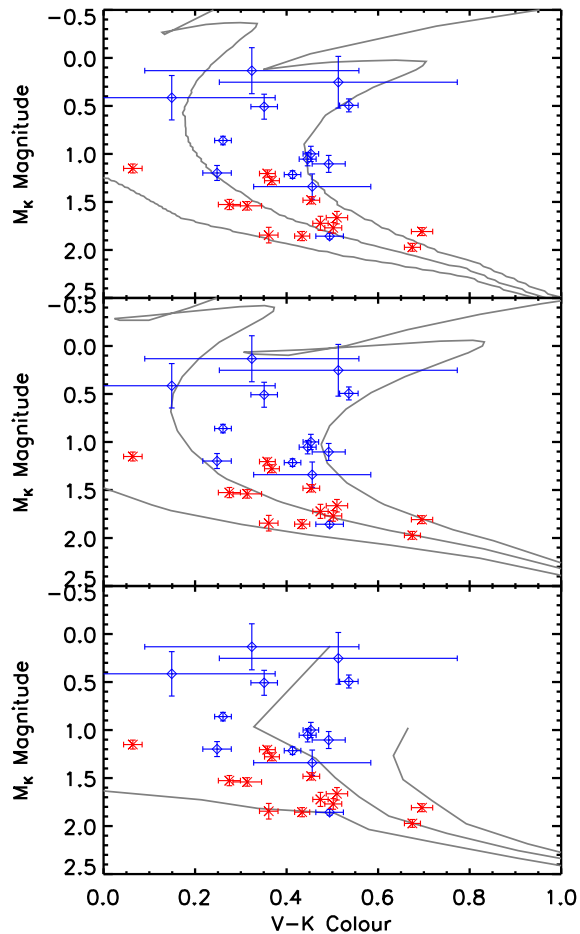}}} 
\caption{A colour-magnitude diagram of the 26 stars discussed within
  this work, plotted alongside three different sets of theoretical
  isochrones at 100, 500, and 800 Myrs. Those targets with new or refined orbits are plotted
  in blue with a diamond symbol (see Table \ref{tab:orbitsample_obs}), and
  those targets for which further measurements are required are
  plotted in red with a cross symbol (see Table \ref{tab:futuresample_obs} and
  \S 6.3). Three different sets of theoretical isochrones are plotted at
ages of 100, 500, and 800 Myrs; (\textit{top}) \citet{Lejeune:2001fq},
(\textit{middle}) \citet{Marigo:2008fy}, and (\textit{bottom})
\citet{Siess:2000tk}.}
\label{fig:cmd}
\end{figure}
\begin{figure}
\resizebox{0.9\hsize}{!}{{\includegraphics{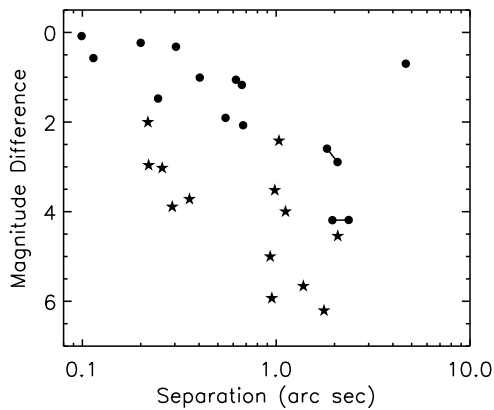}}} 
\caption{The magnitude difference between primary and secondary for
  each binary system within this study as a function of
  angular separation. The targets are divided into two distinct
  subsamples, those with new or refined orbits estimated within this
  study (filled circles), and those for which further measurements are
required before an orbital determination can be attempted (filled
stars). The majority of the systems within the second subsample were
newly resolved as a part of the VAST survey, demonstrating the higher dynamic range
possible with AO imaging. The filled circles connected with a solid
line represent companions within the same hierarchical triple (HIP
44127 and HIP 82321).}
\label{fig:dm}
\end{figure}

The sample of binaries is drawn from the ongoing VAST survey
\citep{DeRosa:2011ci}, an adaptive optics (AO) imaging survey of
A-stars within 75 parsecs, and includes the 26 systems with projected
separations less than 100 AU. The angular separations of the binaries
range from 0$\farcs$094 to 4$\farcs$66, and 11 are newly
resolved. Figure \ref{fig:dm} plots the measured magnitude difference as
a function of separation for the complete sample. Of the 26 systems,
13 have a substantial number of previous measurements, and these
systems comprise the orbit subsample. For the remaining 13
systems, there is insufficient coverage to fit an orbit, and these
binaries comprise the monitoring subsample. Table
\ref{tab1} lists each observed binary in the two subsamples, along
with basic parameters for each such as distance
\citep{vanLeeuwen:2007dc}, \textsl{Tycho2} $V_{\rm T}$-band and
\textsl{2MASS} $K_{\rm s}$-band photometry
\citep{ESA:1997ws,Skrutskie:2006hla}, and spectral type listed within
the \textsl{SIMBAD} database.

For the brightest stars within the sample, the shortest \textsl{2MASS}
exposures saturate, requiring a different method for measuring the
photometry, resulting in significantly larger uncertainties on the
estimated magnitudes \citep{Skrutskie:2006hla}. Therefore for three of these
brighter targets, near-infrared photometry was obtained from
alternative sources \citep{Ducati:2002tr,Morel:1978vm}, and converted
into the \textsl{2MASS} photometric bands using empirical colour
transformations \citep{Carpenter:2001fv}. The distribution of the
sample on the color magnitude diagram (CMD) is plotted in Figure
\ref{fig:cmd}. Given the
rapid evolution of massive stars off the Main Sequence, the position
of an A-star on the CMD provides a method to estimate the age of
the system based on a comparison with theoretical isochrones. The
inferred age of the system from the CMD is combined with the dynamical
system mass from the orbit and system photometry from the literature
to test mass-magnitude relations at the corresponding age.

\section{Observations}
\begin{table*}
\caption{Observing run details with calculated astrometric calibration values}
\begin{tabular}{cccr@{$\pm$}lr@{$\pm$}lc}
Telescope&Date&Filter&\multicolumn{2}{c}{Plate Scale}&\multicolumn{2}{c}{True North}&Proposal ID\\
(Instrument)&&&\multicolumn{2}{c}{(mas/px)}&\multicolumn{2}{c}{($^{\circ}$)}&\\
\hline
\hline
CFHT&05/05/2001 -- 07/05/2001&$H$ continuum&34.87 & 0.14&1.47 & 0.19&2001AF11\\
(KIR)&06/05/2001&$J$ continuum&34.82 & 0.15&1.51 & 0.17&2001AF11\\
&13/06/2008&H$_2$ ($v=1-0$)&34.81 & 0.10&-2.42 & 0.13&2008AC22\\
&14/06/2008&FeII&34.76 & 0.08&-2.40 & 0.11&2008AC22\\
&30/08/2009 -- 01/09/2009&H$_2$ ($v=1-0$)&34.78 & 0.10&-2.40 &
0.20&2009BC06\\
&04/02/2010 -- 05/02/2010&H$_2$ ($v=1-0$)&34.75 & 0.08&-2.37 &
0.11&2010AC14\\
&2011&H$_2$ ($v=1-0$)&34.77 & 0.06&-2.35 & 0.09&2011AC11\\
Gemini N&08/09/2008 -- 14/11/2008&Br$\gamma$&21.28 & 0.12&0.52 & 0.29&GN-2008A-Q-74\\
(NIRI)&19/12/2009&Br$\gamma$&21.28 & 0.12&0.52 & 0.29&GN-2008B-Q-119\\
&25/06/2010 -- 26/08/2010&Br$\gamma$&21.28 & 0.12&0.52 & 0.29&GN-2010A-Q-75\\
Lick&24/07/2008 -- 25/07/2009&Br$\gamma$&74.55 & 1.00&1.68 & 0.50&--\\
(IRCAL)&16/10/2008 -- 17/10/2008&Br$\gamma$&74.55 & 1.00&1.68 & 0.50&--\\
VLT-UT4&30/06/2004&IB218&27.02 & 0.14&0.24 & 0.30&272.D-5068(A)\\
(NACO)&10/01/2005&IB218&27.02 & 0.15&0.26 & 0.30&074.D-0180(A)\\
&08/02/2005&K$_{\rm s}$&27.02 & 0.15&0.26 & 0.30&074.D-0180(A)\\
&08/11/2005&K$_{\rm s}$&26.98 & 0.08&0.21 & 0.20&076.C-0270(A)\\
&10/11/2005&K$_{\rm s}$&26.98 & 0.08&0.21 & 0.20&076.D-0108(A)\\
&06/12/2005&IB218&26.98 & 0.08&0.21 & 0.20&076.D-0108(A)\\
&04/01/2006&K$_{\rm s}$&27.01 & 0.13&0.31 & 0.29&076.D-0108(A)\\
&07/01/2006&IB218&27.01 & 0.13&0.31 & 0.29&076.D-0108(A)\\
&27/04/2006&IB218&27.00 & 0.12&0.09 & 0.26&077.D-0147(A)\\
&20/09/2007&IB218&26.99 & 0.12&-0.23 & 0.21&080.D-0348(A)\\
&20/09/2007&K$_{\rm s}$&26.99 & 0.12&-0.23 & 0.21&080.D-0348(A)\\
&26/09/2007&IB218&26.99 & 0.12&-0.23 & 0.21&080.D-0348(A)\\
&04/11/2007&K$_{\rm s}$&27.00 & 0.10&-0.18 & 0.20&080.D-0348(A)\\
&15/02/2008&IB218&27.01 & 0.14&-0.12 & 0.29&080.D-0348(A)\\
&24/02/2008&K$_{\rm s}$&27.01 & 0.14&-0.12 & 0.29&080.D-0348(A)\\
Palomar&11/04/2008&CH$_4$&25.00&1.00&0.70&1.00&--\\
(PHARO)&12/04/2008&Br$\gamma$&25.00&1.00&0.70&1.00&--\\
&12/07/2008&CH$_4$&25.00&1.00&0.70&1.00&--\\
&13/07/2008&Br$\gamma$&25.00&1.00&0.70&1.00&--\\
\hline
\multicolumn{8}{l}{KIR -- \citep{Doyon:1998vi}}\\
\multicolumn{8}{l}{NIRI (Near InfraRed Imager and Spectrometer) -- \citep{Hodapp:2003ko}}\\
\multicolumn{8}{l}{IRCAL -- \citep{Lloyd:2000we}} \\
\multicolumn{8}{l}{NACO (Nasmyth Adaptive Optics System/
  Near-Infrared Imager and Spectrograph) -- \citep{Lenzen:2003iu,Rousset:2003hh}}\\
\multicolumn{8}{l}{PHARO (Palomar High Angular Resolution Observer) --
  \citep{Hayward:2001kq}}\\
\label{tab:instruments}
\end{tabular}
\end{table*}
With AO systems operating on telescopes ranging in diameter from the
3m Shane to the 8m Gemini and VLT, near-infrared images were obtained
on all targets. In
most cases, the filter used for the observations was a narrow or
broadband filter within the $K$ bandpass, though some images were
taken within the $J$ and $H$ bandpasses. Both the primary and
secondary of the pairs were unsaturated in the AO images, simplifying the astrometry
measurements. Table \ref{tab:instruments} details the key
characteristics of the instruments used to acquire the new AO
observations, with the measured pixel scale and orientation for each
camera. A subset of the observations were obtained from the CFHT and
ESO Science Archive Facilities. One measurement obtained at the
Southern Observatory for Astrophysical Research (SOAR) as a part of
the VAST survey, and used within this study, has been recently
published in \cite{Hartkopf:2012jl}.

\section{Data Analysis}
\subsection{AO image processing}
The AO science images obtained were processed with standard image
reduction steps including dark subtraction, flat fielding,
interpolation over bad pixels and sky subtraction. To align all the
images, the centroid of the bright primary was obtained in each
exposure by fitting a Gaussian to the core of the central point spread
function. For each system resolved within the observations, an
empirical PSF was determined from the radial profile of the
primary, after masking any close companion. The empirical PSF was then
fit to the position and
intensity of both components of the system, providing a measure of the
separation, position angle, and magnitude difference. Uncertainties
within the photometry and astrometry were estimated from the standard
deviations of the photometric and astrometric measurements from each
individual exposure before combination.

To ensure accurate determination of the separation and position angle,
the pixel scale and orientation of the detectors were calibrated based
on observations of the Trapezium cluster, with the exception of data
obtained with IRCAL and PHARO. Depending on the total field-of-view of
the detector, the positions of 20 to 40 Trapezium members were compared
with the coordinates reported in \cite{McCaughrean:1994id}. The
average derived pixel scale and orientation were computed, and the
standard deviation of these values were used as the associated errors;
the results are given in Table \ref{tab:instruments}. For the data
obtained with IRCAL and PHARO, the pixel scale and orientation were
calibrated from binary systems also observed with instruments
calibrated with Trapezium measurements.

\subsection{Orbital determination}
For the 13 binaries with sufficient coverage of the orbit, a fit was
performed for the orbital elements and an estimate of the dynamical
mass was determined. The measurements presented within this study were
combined with previous measurements contained within the Washington
Double Star (WDS; \citealp{Mason:2001ff}) Catalog. These archive
measurements were obtained using a variety of observational
techniques, and date back to the 18$^{\rm th}$ Century. As in some cases
the statistical uncertainties were not provided in the WDS Catalog,
we searched the literature for the formal errors for each individual
measurement, and only separation and position angle values for which
uncertainties could be assigned were included within the fitting
procedure. A detailed listing of the individual measurements used for
the orbital determination will be made available at the Strasbourg
astronomical Data Center (CDS - \texttt{http://cds.u-strasbg.fr}).

Our orbit fitting approach utilises the method presented by
\cite{Hilditch:2001wp}, and demonstrated by an application to
measurements of the T Tau S system \citep{Kohler:2008ek}. This method
is  similar to the grid-based search technique developed by \cite{Hartkopf:1989ez}. At each
epoch of observation $t_i$, the $x_i$,$y_i$ position of the companion
with respect to the primary is measured in the observed tangent
plane. These values are related to the true position of the secondary
in the orbital plane ($x_i^{\prime}$,$y_i^{\prime}$) through the
following equations
\begin{equation}
\begin{split}
x_i&=Ax_i^{\prime}+Fy_i^{\prime}\\
y_i&=Bx_i^{\prime}+Gy_i^{\prime}
\label{eqn:thiele}
\end{split}
\end{equation}
where $A$, $B$, $F$ and $G$ are the orbital Thiele-Innes elements, with
\begin{equation}
\begin{split}
A&=a(\cos\omega\cos\Omega - \sin\omega\sin\Omega\cos i)\\
B&=a(\cos\omega\sin\Omega + \sin\omega\cos\Omega\cos i)\\
F&=a(-\sin\omega\cos\Omega - \cos\omega\sin\Omega\cos i)\\
G&=a(-\sin\omega\sin\Omega + \cos\omega\cos\Omega\cos i)
\label{eqn:thieleb}
\end{split}
\end{equation}
where $a$ is the semi-major axis of the orbit, $\omega$ the longitude
of periastron, $\Omega$ the longitude of the ascending node, and $i$
the inclination -- four of the seven orbital elements. The position of
the companion in the orbital plane ($x_i^{\prime}$,$y_i^{\prime}$) can also be expressed
through the remaining orbital elements ($e$, $P$, $T_0$) as
\begin{equation}
\begin{split}
x_i^{\prime}&=\cos E-e\\
y_i^{\prime}&=\sqrt{1-e^2}\sin E
\label{eqn:xyorbit}
\end{split}
\end{equation}
where $e$ is the eccentricity of the system. The eccentric anomaly
($E$) can be determined from a numerical solution to Kepler's equation 
\begin{equation}
\begin{split}
M&=E-e\sin E\\
&=(2\pi/P)(t_i-T_0)
\end{split}
\end{equation}
where $M$ is the mean anomaly, $P$ the period of the system, and $T_0$
the epoch of periastron passage. At each epoch of observation, the
position of the component in the orbital plane can be defined using
just three of the orbital elements ($e$, $P$, and $T_0$). The orbital
position at each epoch can then be converted into the observed
position using the equations in Equation \ref{eqn:thiele} through a
least-squares determination of the four Thiele-Innes elements.

An initial estimation of the orbital parameters of each system can be
determined through an iterative three-dimensional grid search of $e$,
$P$, and $T_0$. A wider range of parameter values were searched, with
100 linear steps searched over a range of $0\le e < 1$, 500 linear
steps between $1 \le \log(P[yrs]) \le 3$, and $T_0$ initially distributed between the years
$2000.0-(P/2)$ and $2000.0+(P/2)$. At each position within this
three-dimensional grid, the fit orbital positions
($x_i^{\prime}$,$y_i^{\prime}$) were directly
calculated (Eqn. \ref{eqn:xyorbit}), with the four remaining orbital
parameters ($a$, $i$, $\omega$, $\Omega$) estimated from a
least-squares fit to the observed positions using Equations
\ref{eqn:thiele} and \ref{eqn:thieleb}. After the $\chi^2$ statistic
was calculated at each position within the grid, the range of $T_0$
values searched was reduced by a factor of 10 centred on the optimum
value of $T_0$ found within the previous iteration. This process was
repeated until the step size in $T_0$ was reduced to less than one
day. The values for $a$, $i$, $\omega$ and $\Omega$ can be determined
from an inversion of Equation \ref{eqn:thieleb} \citep{Green:1985ve},
\begin{equation}
\begin{split}
\Omega&=\frac{1}{2}\left(\arctan\left(\frac{B-F}{G+A}\right)-\arctan\left(\frac{B+F}{G-A}\right)\right)\\
\omega&=\left(\arctan\left(\frac{B-F}{G+A}\right)-\Omega\right)\\
i&=2\arctan\left[\sqrt{\frac{-\left(B+F\right)\sin\left(\omega+\Omega\right)}{\left(B-F\right)\sin\left(\omega-\Omega\right)}}\right]\\
a&=\frac{B-F}{\sin\left(\omega+\Omega\right)\left(1+\cos\left(i\right)\right)}
\end{split}
\end{equation}
The orbital parameters calculated at each position within the ($P$,
$e$) grid are then used as a starting point for a Levenberg-Marquardt
minimisation to ensure the minimum of the $\chi^2$ distribution is
found. The set of orbital parameters with the minimum $\chi^2$
statistic was then used as the orbit solution for the system.

The shape of the $\chi^2$ distribution in the vicinity of the global
minimum can be used to determine the 1$\sigma$ uncertainties of each
parameter \citep{Press:1992vz}. By perturbing an individual parameter
away from the global minimum, and optimising the remaining parameters, a region
of the $\chi^2$ distribution can be calculated where the $\chi^2$ statistic
is less than $\chi^2_{\rm min}+1$. This region encloses 68\% of the probability
distribution, and is not necessarily symmetric about the minimum
$\chi^2$ value. Our implementation of the orbit fitting method was
tested against four well studied systems \citep{Bonnefoy:2009jc,Dupuy:2009jba,Liu:2008ib}, with the resulting
parameters being within the published 1$\sigma$ uncertainties.

\subsection{Theoretical mass-magnitude relations}
\begin{table}
\caption{Summary of theoretical model grids}
\begin{tabular}{cr@{ $<M<$ }lc}
Grid &\multicolumn{2}{c}{Mass Range} & Metallicity\\
reference&\multicolumn{2}{c}{(M$_{\odot}$)}&($Z$)\\
\hline
\citet{Lejeune:2001fq}&$0.80$&$5.0$&0.02\\
\citet{Marigo:2008fy}&$0.15$&$5.0$&0.02\\
\citet{Siess:2000tk}&$0.10$&$5.0$&0.02\\
\citet{Baraffe:1998ux}&$0.08$&$1.5$&0.02\\
\hline
\label{tab:models}
\end{tabular}
\end{table}

\begin{figure}
\resizebox{0.99\hsize}{!}{{\includegraphics{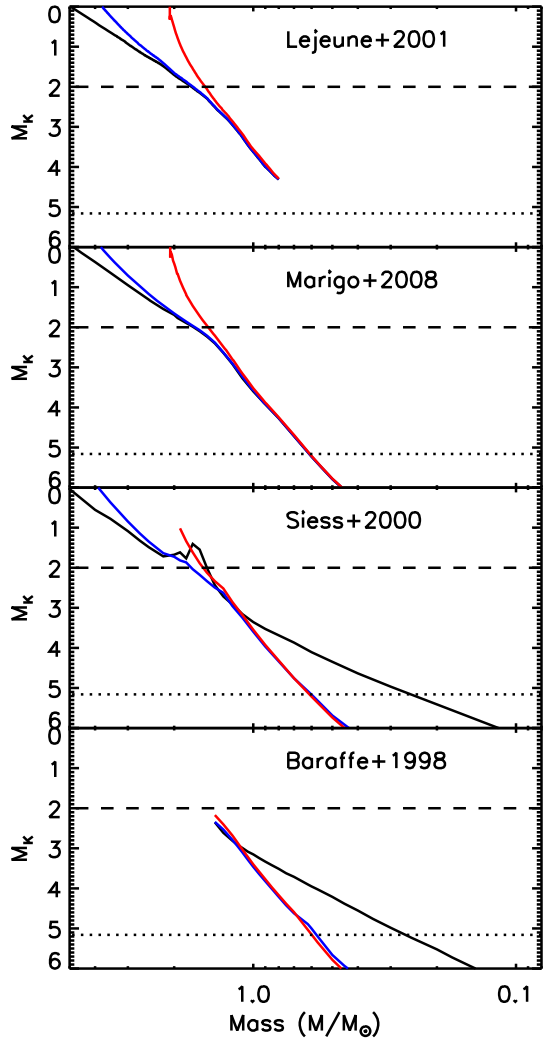}}} 
\caption{Mass-magnitude relations were
  constructed from each of the four model grids. The evolution of the
  mass-magnitude relation is shown for each grid, with the 10 Myr (black line),
  100 Myr (blue line) and 1 Gyr (red line) relations plotted. For
  reference the dashed line indicates the faintest A-type star within
  the sample, and the dotted line indicates the Zero-Age Main Sequence
  magnitude of an M0 star. A-type stars typically have an absolute
  $K$-band magnitude ranging between $M_K=0$  and $M_K=2$, where the
  mass is significantly dependent on the age of the star. Within the
  \citet{Lejeune:2001fq} and \citet{Marigo:2008fy} models, the
  mass-magnitude relation of stars fainter than $M_K \sim 2$ is not
  dependent on the age of the star. The \citet{Siess:2000tk} and
  \citet{Baraffe:1998ux} models include a description of the
  contraction phase of lower-mass stars onto the Main Sequence during
  the early portion of its life.}
\label{fig:massmag}
\end{figure}
\begin{figure}
\resizebox{0.99\hsize}{!}{{\includegraphics{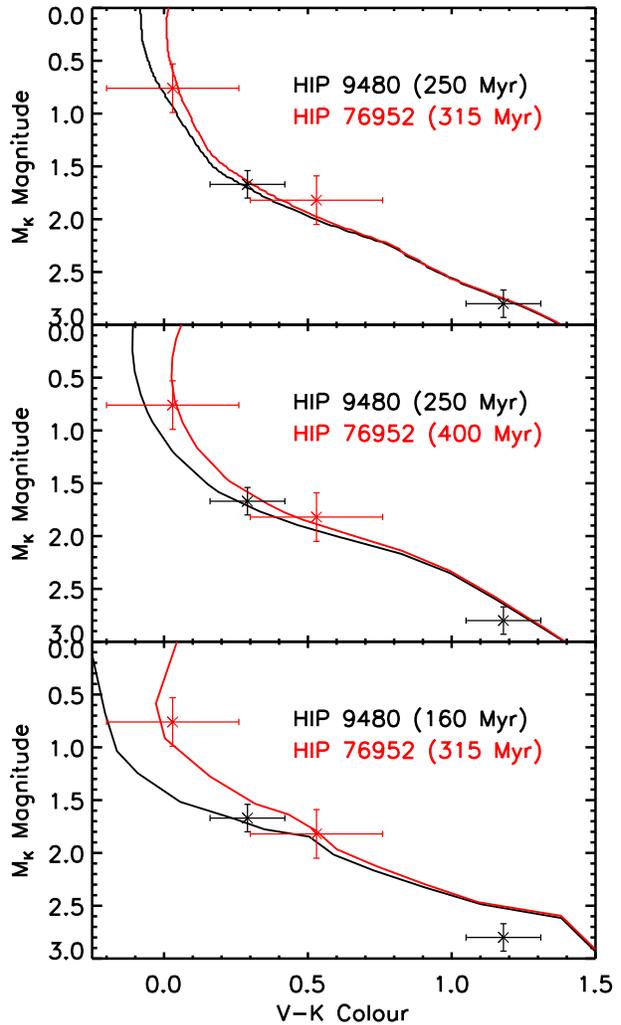}}} 
\caption{The age of each system can be estimated based on the position
of each component on the colour-magnitude diagram (top -
\citet{Lejeune:2001fq}, middle - \citet{Marigo:2008fy}, bottom -
\citet{Siess:2000tk}). This procedure, while demonstrated for only two
systems in this figure, was repeated for all the remaining members of
the orbit subsample. For each system, the age estimated from each of
the three model grids is reported in Table \ref{tab:comparison}.}
\label{fig:hr_double}
\end{figure}
\begin{figure}
\resizebox{0.99\hsize}{!}{{\includegraphics{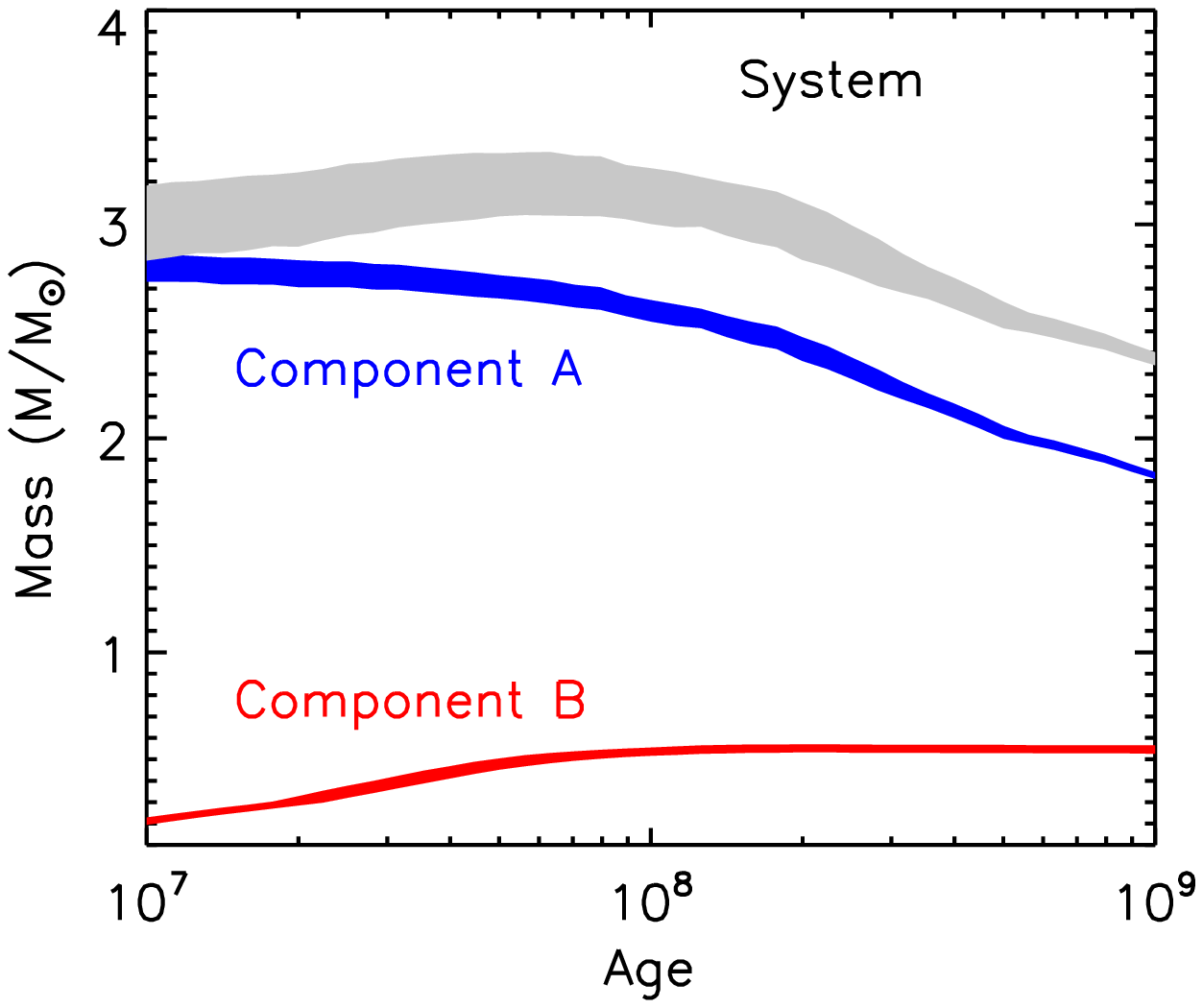}}} 
\caption{The mass-magnitude relation for a hypothetical binary system with $M_K
({\rm A})=1.25\pm0.05$ and $M_K ({\rm B})=5.50\pm0.10$, as a
function of system age. The evolution of the mass-magnitude relation
is shown for both components (A - blue region, B - red region), and the
system as a whole (grey region). The extent of the region in each case
represents the uncertainty in the mass estimate due to the
uncertainties of the magnitudes of each component.}
\label{fig:guide}
\end{figure}

We obtained four different grids of evolutionary models with which
the dynamical system masses estimated from
the fitted orbital parameters were compared. Table \ref{tab:models}
lists the mass range, metallicity, and literature reference for each
of the four grids. The \citet{Lejeune:2001fq},
\citet{Marigo:2008fy}, and \citet{Siess:2000tk} grids
covers a significant portion of the lifespan of a typical A-type star,
and as such we applied a maximum age cut-off at 1 Gyr. In addition to
these grids, models from \citet{Baraffe:1998ux} were obtained in
order to study a pair of lower-mass companions presented in \S
6.1.2. Each grid was converted into the photometric systems used
within this study - \textsl{Tycho} $V$ and \textsl{2MASS} $K_{S}$
\citep{Carpenter:2001fv}, before producing a high resolution
($dM/M_{\odot}=0.001$) mass-magnitude relation, created through cubic
interpolation of the grid data, as shown in Figure \ref{fig:massmag}.

The absolute $V$- and $K$-band magnitudes were calculated from the
$V$-band magnitude differences obtained from the literature, and the $K$-band magnitude differences presented within
this study (Tables \ref{tab:orbitsample_obs} and \ref{tab:futuresample_obs}). The individual component
magnitudes and $V-K$ colour for each system are presented in Table
\ref{tab:components}. Within the A-type star mass range, the
mass-magnitude relations significantly change as a function of the age
of the system due to the rapid evolution of A-type stars across the
CMD. The age of each system is therefore estimated, based on
the position of the primary on the CMD (Figure \ref{fig:hr_double}), before a mass for each
component is estimated from the mass-magnitude relations. The
estimated masses of each component are summed to produce an estimation
of the system mass, hereafter called the theoretical system mass.

To demonstrate the analysis procedure, a
hypothetical binary system of magnitudes $M_K ({\rm A})=1.25$, $M_K
({\rm B})=5.50$ was used to construct the mass-age relation for each
component, and their corresponding sum (Figure \ref{fig:guide}). The
evolution of the mass-magnitude relation as a function of age can be
then visualised as a continuous function for both components within the system. For
this example a set of models were used which includes the pre-Main
Sequence (PMS) contraction phase of lower-mass stars, as demonstrated
by the increase in the derived mass as a function of age for the
companion. The two mass-age curves can then be summed
to produce a mass-age relation for the system in question. Using the
age estimated for the system based upon its position on the CMD, a
theoretical system mass can be estimated and compared with the
dynamical system mass determined from the orbital
elements.

\section{Results}
\subsection{Astrometric results}
\begin{table*}
\caption{Measured binary position angle and separation for previously
  resolved systems}
\begin{tabular}{lccccccr@{$\pm$}lr@{$\pm$}lr@{$\pm$}l}
HIP&Comp.&WDS&Discoverer&Previous&Instrument&Epoch&\multicolumn{2}{c}{$\theta$}&\multicolumn{2}{c}{$\rho$}&\multicolumn{2}{c}{$\Delta K$}\\
&&Designation&Designation&Orbit&&&\multicolumn{2}{c}{(deg)}&\multicolumn{2}{c}{(arcsec)}&\multicolumn{2}{c}{(mag)}\\
\hline
\hline
5300&AB&01078-4129&RST3352&\cite{Soderhjelm:1999wc}&NACO&2005.02&15.73&0.76&0.089&0.001&0.507&0.020\\
 &&&&&NACO&2007.72&126.69&0.38&0.114&0.001&*0.573&0.018\\
9480&A-B&02020+7054&BU 513&\cite{Mason:1999cd}&IRCAL&2008.79&293.97&0.75&0.679&0.011&1.110&0.056\\
&&&&&KIR&2009.67&297.36&0.22&0.666&0.002&1.172&0.013\\
&&&&&NIRI&2010.65&302.11&0.29&0.644&0.004&*1.137&0.007\\
11569&Aa,Ab&02291+6724&CHR 6&\cite{Drummond:2003fl}&IRCAL&2008.79&43.50&0.80&0.575&0.006&*1.906&0.026\\
&&&&&KIR&2010.10&41.52&2.61&0.602&0.023&1.771&0.301\\
17954&AB&03503+2535& STT 65&\cite{Docobo:2007df}&NIRI&2008.87&194.76&0.31&0.200&0.001&*0.232&0.005\\
28614&AB&06024+0939&A 2715&\cite{Muterspaugh:2008fg}&NIRI&2009.96&21.95&0.29&0.403&0.002&*1.008&0.006\\
&&&&&KIR&2011.31&21.31&1.06&0.397&0.006&0.896&0.120\\
36850&AB&07346+3153&STF1110&\cite{Docobo:1985un}&KIR&2010.09&57.46&0.14&4.663&0.013&*0.698&0.050\\
44127&BC&08592+4803&HU 628&\cite{Eggen:1967bg}&PHARO&2008.28&242.01&1.04&0.433&0.017&*0.005&0.031\\
&&&&&KIR&2010.10&228.98&0.27&0.568&0.003&0.027&0.023\\
&&&&&KIR&2011.30&223.37&0.31&0.656&0.003&-0.011&0.039\\
47479&AB&09407-5759&B 780&\cite{Soderhjelm:1999wc}&NACO&2008.12&300.22&0.58&0.099&0.001&*0.082&0.017\\
76952&AB&15427+2618&STF1967&\cite{Muterspaugh:2010cl}&PHARO&2008.28&112.65&1.07&0.683&0.028&\multicolumn{2}{c}{-}\\
&&&&&KIR&2011.31&111.37&0.33&0.620&0.005&*1.056&0.030\\
77660&AB&15513-0305&CHR 51&\cite{Docobo:2010hf}&KIR&2001.34&82.72&2.42&0.188&0.018&\multicolumn{2}{c}{-}\\
&&&&&KIR&2001.34&81.97&1.59&0.197&0.008&\multicolumn{2}{c}{-}\\
&&&&&KIR&2001.35&83.38&3.67&0.185&0.016&\multicolumn{2}{c}{-}\\
&&&&&NACO&2004.50&71.87&0.31&0.247&0.001&*1.487&0.006\\
&&&&&NACO&2006.32&69.74&0.42&0.274&0.003&1.472&0.097\\
&&&&&KIR&2010.10&66.69&0.47&0.330&0.004&1.483&0.058\\
&&&&&KIR&2011.30&64.21&2.57&0.340&0.014&1.455&0.194\\
80628&Ab,Ab&16278-0822&RST3949&-&PHARO&2008.28&22.56&1.05&0.674&0.027&*2.072&0.066\\
&&&&&IRCAL&2008.56&23.53&0.71&0.685&0.009&1.978&0.057\\
&&&&&KIR&2011.31&34.37&0.29&0.771&0.003&2.237&0.038\\
82321&BC&16492+4559&A 1866&\cite{Popovic:1969vp}&PHARO&2008.53&247.33&1.03&0.286&0.012&\multicolumn{2}{c}{-}\\
&&&&&KIR&2011.30&260.51&0.53&0.301&0.005&*0.299&0.044\\
93506&AB&19026-2953&HDO 150&\cite{Mason:1999cd}&KIR&2008.45&31.05&2.31&0.194&0.010&\multicolumn{2}{c}{-}\\
&&&&&KIR&2011.31&285.59&0.71&0.304&0.004&*0.320&0.043\\
\hline
\multicolumn{13}{l}{* - Photometry measurement used to determine
  the magnitude of each component in Table \ref{tab:components}.}
\label{tab:orbitsample_obs}
\end{tabular}
\end{table*}
\begin{table*}
\caption{Measured binary position angle and separation for potential
  orbital monitoring targets}
\begin{tabular}{lccr@{$\pm$}lr@{$\pm$}lr@{$\pm$}lr@{$\pm$}l}
HIP&Instrument&Epoch&\multicolumn{2}{c}{$\theta$}&\multicolumn{2}{c}{$\rho$}&\multicolumn{2}{c}{$\Delta
  K$}&\multicolumn{2}{c}{Projected}\\
&&&\multicolumn{2}{c}{(deg)}&\multicolumn{2}{c}{(arcsec)}&\multicolumn{2}{c}{(mag)}&\multicolumn{2}{c}{Separation
(AU)}\\
\hline
\hline
128&NIRI&2008.72&80.64&0.29&0.983&0.006&*3.521&0.021&69.59&1.72\\
&IRCAL&2008.79&80.73&0.52&0.975&0.013&3.355&0.037&68.98&1.90\\
2381&NACO&2005.93&279.51&0.26&1.765&0.006&*6.206&0.088&93.75&1.48\\
&NACO&2007.73&279.13&0.22&1.767&0.008&6.229&0.055&93.86&1.51\\
&NIRI&2008.79&278.54&0.29&1.759&0.013&7.401&0.711&93.39&1.59\\
2852&NIRI&2008.79&260.55&0.30&0.931&0.006&*5.001&0.056&45.51&0.82\\
5310&NIRI&2008.79&174.99&0.30&0.357&0.002&*3.719&0.023&16.88&0.62\\
18217&IRCAL&2008.79&64.78&0.65&1.037&0.012&2.301&0.029&52.37&1.24\\
&NIRI&2008.87&64.96&0.29&1.033&0.006&*2.41&0.010&52.13&1.12\\
29852&NACO&2005.85&210.76&0.22&0.218&0.001&*2.003&0.013&13.49&0.21\\
&NACO&2005.86&210.80&0.21&0.217&0.001&1.983&0.012&13.44&0.21\\
51384&PHARO&2008.18&212.43&1.05&2.078&0.084&*4.543&0.169&84.44&3.66\\
65241&NACO&2005.10&196.95&0.62&0.328&0.003&3.061&0.231&21.65&0.50\\
&NACO&2008.15&212.75&0.57&0.258&0.003&*3.025&0.097&17.01&0.42\\
66223&PHARO&2008.53&187.71&1.02&1.381&0.056&*5.660&0.105&96.42&4.73\\
103298&NIRI&2008.69&115.72&0.30&0.220&0.001&*2.962&0.007&13.29&0.25\\
109667&NIRI&2008.69&285.17&0.30&1.117&0.007&*3.997&0.011&70.88&2.38\\
&KIR&2009.66&284.66&0.25&1.108&0.006&4.051&0.073&70.28&2.35\\
&NIRI&2010.48&284.38&0.29&1.101&0.007&4.044&0.103&69.84&2.34\\
110787&NIRI&2008.71&211.11&0.35&0.291&0.002&*3.891&0.043&18.38&0.29\\
116611&NIRI&2008.75&173.11&0.32&0.950&0.006&*5.930&0.089&67.78&1.38\\
&NIRI&2010.48&172.11&0.34&0.943&0.006&5.794&0.076&67.31&1.34\\
\hline
\multicolumn{11}{l}{* - Photometry measurement used to determine
  the magnitude of each component in Table \ref{tab:components}.}
\label{tab:futuresample_obs}
\end{tabular}
\end{table*}
\begin{table*}
\caption{Derived parameters of individual components}
\begin{tabular}{lcr@{$\pm$}lr@{$\pm$}lr@{$\pm$}lr@{$\pm$}lr@{$\pm$}lccc}
HIP&Comp.&\multicolumn{2}{c}{$V_{\rm T}$}&\multicolumn{2}{c}{$K_{\rm
    S}$}&\multicolumn{2}{c}{$V-K$}&\multicolumn{2}{c}{${\rm
    M}_V$}&\multicolumn{2}{c}{${\rm M}_K$}&Additional&SB&Reference\\
&&\multicolumn{2}{c}{(mag)}&\multicolumn{2}{c}{(mag)}&\multicolumn{2}{c}{}&\multicolumn{2}{c}{(mag)}&\multicolumn{2}{c}{(mag)}&Components&Type&\\
\hline
\hline
\multicolumn{13}{c}{Orbit Subsample}\\
\hline
5300&A&5.51&0.02&5.28&0.02&0.23&0.03&1.73&0.08&1.50&0.08\\
&B&6.85&0.08&5.85&0.02&0.99&0.08&3.07&0.11&2.08&0.08\\
9480$^{\dagger}$&A&4.69&0.01&4.40&0.13&0.29&0.13&1.95&0.03&1.67&0.13\\
&B&6.72&0.01&5.54&0.13&1.18&0.13&3.98&0.04&2.80&0.13&\\
11569 $^{\ddagger}$&Aa&4.67&0.01&4.79&0.03&-0.12&0.03&1.62&0.07&1.74&0.08\\
&Ab&8.65&0.16&6.71&0.03&1.94&0.16&5.60&0.17&3.66&0.08\\
17954&A&5.81&0.04&5.46&0.02&0.35&0.04&2.05&0.08&1.70&0.07\\
&B&6.26&0.06&5.69&0.02&0.57&0.06&2.50&0.09&1.93&0.07\\
28614&A&4.32&0.01&4.00&0.26&0.33&0.26&0.94&0.07&0.61&0.27&Aa, Ab&SB1&\cite{Fekel:2002ga}\\
&B&6.22&0.02&5.01&0.26&1.22&0.26&2.84&0.07&1.62&0.27&Ba, Bb&SB2&\cite{Fekel:2002ga}\\
36850&A&1.98&0.02&1.93&0.03&0.05&0.04&1.02&0.13&0.97&0.13&Aa, Ab&SB1&\cite{VinterHansen:1940ws}\\
&B&2.88&0.03&2.63&0.04&0.25&0.05&1.92&0.13&1.66&0.13&Ba, Bb&SB1&\cite{VinterHansen:1940ws}\\
44127&A&3.16&0.02&2.71&0.03&0.45&0.04&2.35&0.02&1.90&0.03\\
&B&10.88&0.11&6.90&0.04&3.98&0.12&10.07&0.11&6.09&0.04\\
&C&11.08&0.12&6.90&0.04&4.18&0.12&10.27&0.12&6.09&0.04\\
47479&A&5.82&0.01&5.51&0.02&0.31&0.02&1.51&0.07&1.21&0.07\\
&B&6.45&0.01&5.59&0.02&0.85&0.03&2.14&0.07&1.29&0.07\\
76952&A&4.05&0.01&4.02&0.23&0.03&0.23&0.80&0.05&0.76&0.23\\
&B&5.60&0.02&5.07&0.23&0.53&0.23&2.35&0.05&1.82&0.23\\
77660&A&5.21&0.01&4.94&0.02&0.26&0.02&1.72&0.04&1.46&0.04\\
&B&7.82&0.04&6.43&0.02&1.38&0.04&4.33&0.05&2.95&0.04\\
80628&Aa&4.68&0.01&4.32&0.04&0.37&0.04&1.62&0.08&1.25&0.09&Aa1, Aa2&SB2&\cite{Gutmann:1965tp}\\
&Ab&8.80&0.10&6.39&0.07&2.41&0.12&5.74&0.13&3.33&0.11\\
82321&A&4.87&0.01&4.73&0.02&0.14&0.02&1.16&0.04&1.02&0.04\\
&B&9.19&0.11&7.33&0.03&1.86&0.11&5.47&0.12&3.62&0.05\\
&C&9.29&0.11&7.63&0.04&1.66&0.12&5.57&0.12&3.92&0.05\\
93506&A&3.27&0.01&2.90&0.23&0.37&0.23&1.11&0.05&0.74&0.24\\
&B&3.48&0.01&3.22&0.24&0.26&0.24&1.32&0.05&1.06&0.24\\
\hline
\multicolumn{13}{c}{Monitoring Subsample}\\
\hline
128&AB&\multicolumn{2}{c}{--}&6.06&0.02&\multicolumn{2}{c}{--}&\multicolumn{2}{c}{--}&1.81&0.06&
A, B&SB1&\cite{Carquillat:2003di}\\
&C&\multicolumn{2}{c}{--}&9.58&0.03&\multicolumn{2}{c}{--}&\multicolumn{2}{c}{--}&5.33&0.06\\
2381&A&\multicolumn{2}{c}{--}&4.83&0.02&\multicolumn{2}{c}{--}&\multicolumn{2}{c}{--}&1.21&0.04\\
&B&\multicolumn{2}{c}{--}&11.04&0.09&\multicolumn{2}{c}{--}&\multicolumn{2}{c}{--}&7.41&0.10\\
2852&A&\multicolumn{2}{c}{--}&5.43&0.02&\multicolumn{2}{c}{--}&\multicolumn{2}{c}{--}&1.98&0.04\\
&B&\multicolumn{2}{c}{--}&10.43&0.06&\multicolumn{2}{c}{--}&\multicolumn{2}{c}{--}&6.98&0.07\\
5310&A&\multicolumn{2}{c}{--}&5.25&0.02&\multicolumn{2}{c}{--}&\multicolumn{2}{c}{--}&1.88&0.08\\
&B&\multicolumn{2}{c}{--}&8.97&0.03&\multicolumn{2}{c}{--}&\multicolumn{2}{c}{--}&5.60&0.08\\
18217&A&\multicolumn{2}{c}{--}&5.48&0.02&\multicolumn{2}{c}{--}&\multicolumn{2}{c}{--}&1.97&0.05\\
&B&\multicolumn{2}{c}{--}&7.90&0.02&\multicolumn{2}{c}{--}&\multicolumn{2}{c}{--}&4.38&0.05\\
29852&A&\multicolumn{2}{c}{--}&5.60&0.02&\multicolumn{2}{c}{--}&\multicolumn{2}{c}{--}&1.64&0.04\\
&B&\multicolumn{2}{c}{--}&7.60&0.02&\multicolumn{2}{c}{--}&\multicolumn{2}{c}{--}&3.64&0.04\\
51384&A&\multicolumn{2}{c}{--}&4.87&0.02&\multicolumn{2}{c}{--}&\multicolumn{2}{c}{--}&1.82&0.04\\
&B&\multicolumn{2}{c}{--}&9.41&0.17&\multicolumn{2}{c}{--}&\multicolumn{2}{c}{--}&6.37&0.17\\
65241&A&\multicolumn{2}{c}{--}&5.69&0.02&\multicolumn{2}{c}{--}&\multicolumn{2}{c}{--}&1.59&0.05\\
&B&\multicolumn{2}{c}{--}&8.71&0.09&\multicolumn{2}{c}{--}&\multicolumn{2}{c}{--}&4.62&0.11\\
66223&Aa&\multicolumn{2}{c}{--}&5.89&0.02&\multicolumn{2}{c}{--}&\multicolumn{2}{c}{--}&1.67&0.06\\
&Ab&\multicolumn{2}{c}{--}&11.55&0.11&\multicolumn{2}{c}{--}&\multicolumn{2}{c}{--}&7.33&0.12\\
103298&Aa&\multicolumn{2}{c}{--}&5.26&0.02&\multicolumn{2}{c}{--}&\multicolumn{2}{c}{--}&1.35&0.04\\
&Ab&\multicolumn{2}{c}{--}&8.22&0.02&\multicolumn{2}{c}{--}&\multicolumn{2}{c}{--}&4.31&0.04\\
109667&A&\multicolumn{2}{c}{--}&5.76&0.02&\multicolumn{2}{c}{--}&\multicolumn{2}{c}{--}&1.75&0.07\\
&B&\multicolumn{2}{c}{--}&9.76&0.02&\multicolumn{2}{c}{--}&\multicolumn{2}{c}{--}&5.75&0.07\\
110787&A&\multicolumn{2}{c}{--}&5.57&0.03&\multicolumn{2}{c}{--}&\multicolumn{2}{c}{--}&1.57&0.04&\\
&B&\multicolumn{2}{c}{--}&9.46&0.05&\multicolumn{2}{c}{--}&\multicolumn{2}{c}{--}&5.46&0.06\\
116611&Aa&\multicolumn{2}{c}{--}&5.42&0.02&\multicolumn{2}{c}{--}&\multicolumn{2}{c}{--}&1.16&0.05&Aa1,
Aa2&SB1&\cite{Rucinski:2005gu}\\
&Ab&\multicolumn{2}{c}{--}&11.35&0.09&\multicolumn{2}{c}{--}&\multicolumn{2}{c}{--}&7.09&0.10\\
\hline
\multicolumn{15}{l}{$^{\dagger}$ - The spectroscopic binary reported
  by \cite{Abt:1965fz} is the system resolved within the AO data.}\\
\multicolumn{15}{l}{$^{\ddagger}$ - The individual component
  magnitudes may  be significantly biased due to the presence of additional
  companions within the resolution limit of the}\\
\multicolumn{15}{l}{\textsl{Tycho2} and
  \textsl{2MASS} observations.}
\label{tab:components}
\end{tabular}
\end{table*}
\begin{table}
\caption{Visual magnitude differences for a subset of the sample}
\begin{center}
\begin{tabular}{lr@{$\pm$}ll}
HIP&\multicolumn{2}{c}{$\Delta V$}&Reference\\
\hline
\hline
5300&1.34&0.10*&\citet{Horch:2001fa}\\
9480&2.03&0.01&\citet{Fabricius:2000um}\\
11569&3.98&0.16&\citet{Christou:2006ip}\\
17954&0.45&0.10*&\citet{Horch:2004to}\\
28614&1.90&0.02&\citet{Fabricius:2000um}\\
36850&0.90&0.03&\citet{Worley:1969ih}\\
44127 A,BC&7.06&0.10*&\citet{Baize:1989ur}\\
44127 BC&0.20&0.10*&\citet{Mason:2001ff}\\
47479&0.63&0.02&\citet{Mason:2001ff}\\
76952&1.55&0.02&\citet{Fabricius:2000um}\\
77660&2.61&0.04&\citet{Docobo:2010hf}\\
80628&4.12&0.10*&\citet{Mason:2001ff}\\
82321 A,BC&3.61&0.10*&\citet{Mason:2001ff}\\
83231 BC&0.10&0.10*&\citet{Mason:2001ff}\\
93506&0.21&0.02&\citet{Fabricius:2000um}\\
\hline
\multicolumn{4}{l}{* - For $\Delta$V measurements without
  uncertainties, 0.10 is used}
\label{tab6}
\end{tabular}
\end{center}
\end{table}
The astrometry and photometry measurements of the two subsamples are
given in Tables \ref{tab:orbitsample_obs} and \ref{tab:futuresample_obs}. Both tables
contain the \textsl{Hipparcos} designation of the primary, the
components of the system under investigation, the
instrument and epoch of observation, and
the measured astrometric values with corresponding
uncertainties. For the orbit subsample, the WDS designation and
discoverer code are also listed for reference.

Combining the $K$-band magnitudes of the sample with the $\Delta K$
values reported in Tables \ref{tab:orbitsample_obs} and
\ref{tab:futuresample_obs}, and the \textsl{Hipparcos} parallax, 
allowed for an estimation of the $K$-band apparent and absolute
magnitudes of the resolved components (Table \ref{tab:components}). For
systems with $\Delta V$ measurements reported within the literature
(Table \ref{tab6}), the corresponding estimated $V$-band apparent and
absolute magnitudes, and $V-K$ colours for each resolved component are
reported. Seven members of the overall sample are hierarchical systems with at least
one of the components resolved within our dataset consisting of
multiple sub-components, indicated in Table \ref{tab:components}. An example of
this is HIP 128; our AO data are able to resolve a previously-unknown
binary companion within this system (HIP 128 C at $\sim 1\farcs0$), but
are of insufficient angular resolution to resolve the previously-known
spectroscopic component HIP 128 B. Without an estimate of the $\Delta
V$ or $\Delta K$ between HIP 128 A and HIP 128 B, the individual
magnitudes cannot be estimated and therefore only the blended
magnitudes of the two components are listed.

\subsection{Orbital elements and dynamical masses}
\begin{table*}
\caption{Estimated orbital parameters and corresponding 1$\sigma$ uncertainties}
\begin{tabular}{lcr@{$^+_-$}lr@{$^+_-$}lr@{$^+_-$}lr@{$^+_-$}lr@{$^+_-$}lr@{$^+_-$}lr@{$^+_-$}lr@{$\pm$}l}
HIP&N.&\multicolumn{2}{c}{Period}&\multicolumn{2}{c}{Semi-major}&\multicolumn{2}{c}{Inclination}&\multicolumn{2}{c}{Longitude}&\multicolumn{2}{c}{Epoch of}&\multicolumn{2}{c}{Eccentricity}&\multicolumn{2}{c}{Longitude}&\multicolumn{2}{c}{System}\\
&Meas.&\multicolumn{2}{c}{}&\multicolumn{2}{c}{axis}&\multicolumn{2}{c}{}&\multicolumn{2}{c}{of Node}&\multicolumn{2}{c}{Periastron}&\multicolumn{2}{c}{}&\multicolumn{2}{c}{of Periastron}&\multicolumn{2}{c}{Mass}\\
&&\multicolumn{2}{c}{$P$ (yrs)}&\multicolumn{2}{c}{$a$ ($\arcsec$)}&\multicolumn{2}{c}{$i$ ($^{\circ}$)}&\multicolumn{2}{c}{$\Omega$ ($^{\circ}$)}&\multicolumn{2}{c}{$T_0$ (yrs)}&\multicolumn{2}{c}{$e$}&\multicolumn{2}{c}{$\omega$ ($^{\circ}$)}&\multicolumn{2}{c}{(M$_{\odot}$)}\\
\hline
\hline
5300&19&28.36&$^{0.04}_{0.04}$&0.2396&$^{0.0005}_{0.0005}$&65.3&$^{0.3}_{0.3}$&142.7&$^{0.3}_{0.3}$&2007.94&$^{0.02}_{0.02}$&0.424&$^{0.002}_{0.002}$&333.9&$^{0.4}_{0.3}$&3.16&0.34\\
9480&66&61.14&$^{0.05}_{0.05}$&0.614&$^{0.002}_{0.002}$&16.7&$^{0.9}_{0.9}$&48.2&$^{3.3}_{3.1}$&1964.35&$^{0.09}_{0.09}$&0.355&$^{0.001}_{0.001}$&19.5&$^{3.5}_{3.7}$&2.72&0.13\\
11569&11&50.2&$^{1.0}_{1.0}$&0.429&$^{0.007}_{0.007}$&149.0&$^{1.7}_{1.6}$&180.0&$^{2.7}_{2.8}$&1993.24&$^{0.08}_{0.08}$&0.642&$^{0.009}_{0.009}$&331.3&$^{2.6}_{2.7}$&2.12&0.25\\
17954&44&61.2&$^{0.1}_{0.1}$&0.442&$^{0.002}_{0.002}$&84.7&$^{0.1}_{0.1}$&26.4&$^{0.2}_{0.2}$&1998.30&$^{0.08}_{0.08}$&0.628&$^{0.002}_{0.002}$&340.3&$^{0.7}_{0.7}$&4.15&0.39\\
28614&69&18.641&$^{0.008}_{0.008}$&0.2741&$^{0.0001}_{0.0001}$&96.59&$^{0.02}_{0.02}$&24.76&$^{0.02}_{0.02}$&2003.742&$^{0.004}_{0.004}$&0.744&$^{0.001}_{0.001}$&217.08&$^{0.08}_{0.08}$&6.36&0.62\\
36850&207&466.8&$^{6.3}_{6.1}$&6.78&$^{0.05}_{0.05}$&113.56&$^{0.09}_{0.09}$&41.2&$^{0.1}_{0.1}$&1957.3&$^{0.3}_{0.3}$&0.333&$^{0.007}_{0.006}$&249.3&$^{0.6}_{0.5}$&5.42&0.97\\
44127&9&39.4&$^{0.1}_{0.1}$&0.70&$^{0.01}_{0.01}$&111.6&$^{1.1}_{1.1}$&24.5&$^{1.2}_{1.3}$&1999.1&$^{0.6}_{0.7}$&0.35&$^{0.02}_{0.02}$&354.2&$^{4.4}_{4.7}$&0.68&0.04\\
47479&5&10.74&$^{0.04}_{0.04}$&0.1207&$^{0.0006}_{0.0006}$&128.4&$^{1.5}_{1.4}$&87.4&$^{0.5}_{0.5}$&2005.89&$^{0.03}_{0.04}$&0.365&$^{0.009}_{0.009}$&18.6&$^{1.1}_{1.2}$&5.83&0.53\\
76952&104&91.2&$^{0.4}_{0.4}$&0.729&$^{0.006}_{0.006}$&94.45&$^{0.07}_{0.07}$&111.75&$^{0.09}_{0.09}$&1931.6&$^{0.3}_{0.3}$&0.48&$^{0.01}_{0.01}$&103.8&$^{0.6}_{0.5}$&4.19&0.30\\
80628&8&82.8&$^{1.5}_{1.3}$&0.79&$^{0.03}_{0.02}$&31.2&$^{5.4}_{6.0}$&86.8&$^{8.1}_{5.6}$&1994.1&$^{0.9}_{1.1}$&0.45&$^{0.03}_{0.03}$&177.9&$^{6.8}_{7.9}$&4.99&0.75\\
82321&12&56.4&$^{0.3}_{0.3}$&0.279&$^{0.005}_{0.005}$&37.4&$^{2.6}_{2.8}$&57.5&$^{5.9}_{6.1}$&1991.2&$^{2.3}_{2.1}$&0.13&$^{0.02}_{0.02}$&69.4&$^{8.0}_{7.5}$&1.16&0.09\\
93506&42&21.00&$^{0.01}_{0.01}$&0.489&$^{0.001}_{0.001}$&111.1&$^{0.1}_{0.1}$&74.0&$^{0.1}_{0.1}$&2005.99&$^{0.03}_{0.03}$&0.211&$^{0.001}_{0.001}$&7.2&$^{0.6}_{0.6}$&5.26&0.37\\
\hline
\label{tab:orbits}
\end{tabular}
\end{table*}
\begin{table}
\begin{center}
\caption{Previous and revised dynamical system masses}
\begin{tabular}{lr@{$\pm$}lr@{$\pm$}l}
HIP&\multicolumn{2}{c}{Old Estimate}&\multicolumn{2}{c}{New Estimate}\\
&\multicolumn{2}{c}{(M$_{\odot}$)}&\multicolumn{2}{c}{(M$_{\odot}$)}\\
\hline
\hline
5300&\multicolumn{2}{c}{3.45}&3.16&0.34\\
9480&2.98&0.39&2.72&0.13\\
11569&2.12&0.33&2.12&0.25\\
17954&3.37&0.35&4.15&0.39\\
28614&6.33&0.62&6.36&0.62\\
36850&\multicolumn{2}{c}{5.51}&5.42&0.97\\
44127&0.61&0.03&0.68&0.04\\
47479&\multicolumn{2}{c}{7.50}&5.83&0.53\\
76952&4.14&0.28&4.19&0.30\\
80628&\multicolumn{2}{c}{-}&4.99&0.75\\
82321&\multicolumn{2}{c}{1.00}&1.16&0.09\\
93506&5.20&0.36&5.26&0.37\\
\hline
\label{tab:old_new_mass}
\end{tabular}
\end{center}
\end{table}
\begin{figure*}
\resizebox{0.99\hsize}{!}{{\includegraphics{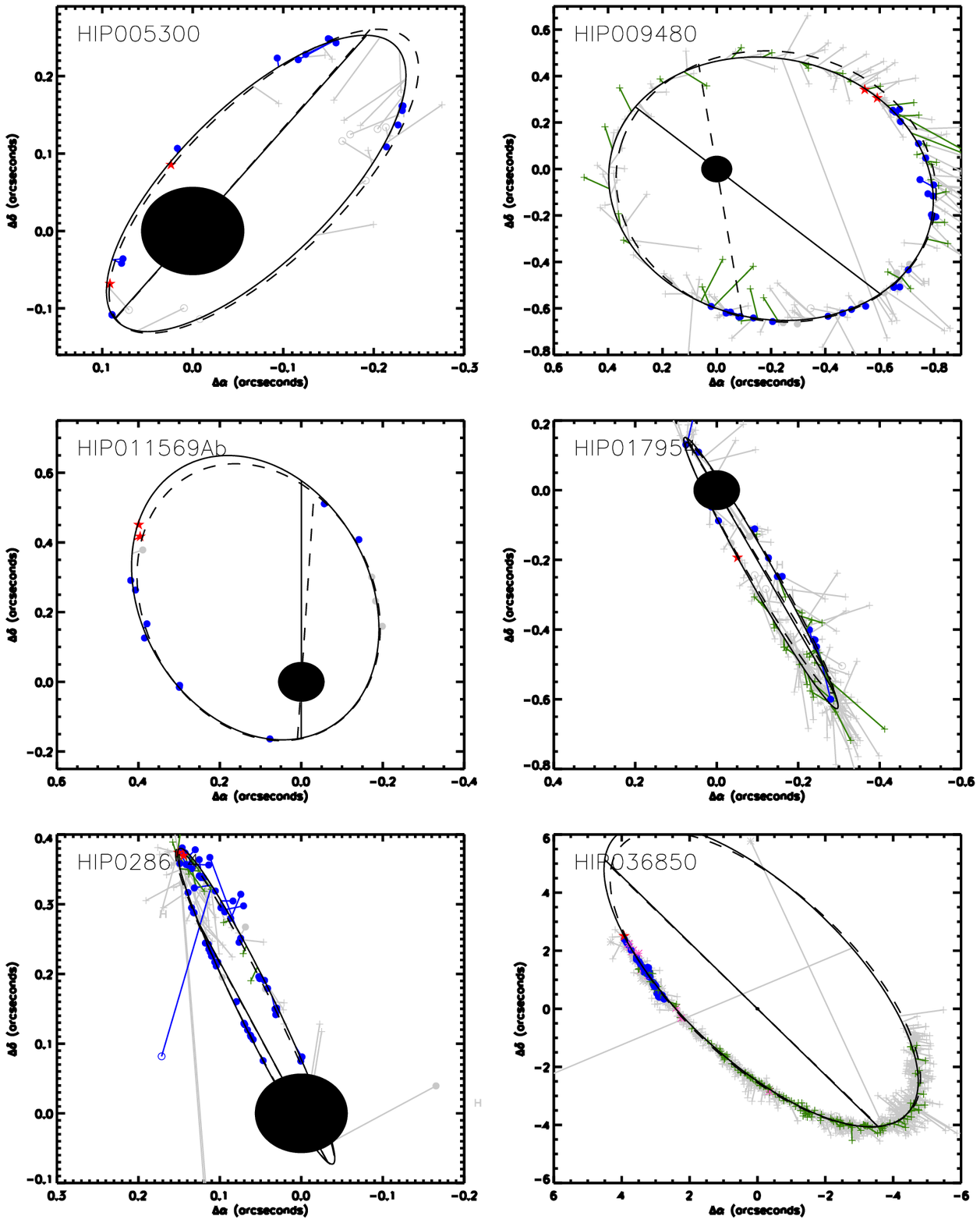}}} 
\caption{Combining our high resolution observations with historical
  measurements, refined orbits for 6 binary systems are plotted. The
  previous orbital fit, obtained from the Sixth Orbit Catalog, is
  plotted for reference with a dashed line. Each plot uses a similar
  symbol scheme to the Sixth Orbit Catalog; open blue circles representing
  eyepiece interferometry, filled blue circles speckle
  interferometry, green crosses micrometrical observations,
and violet asterisks photographic measurements. Our high resolution
observations presented within this study are plotted as filled red
stars. For each measurement, the corresponding O-C line is plotted,
showing the difference between expected and actual position within
the orbital path. Symbols in grey represent those measurements
presented without formal errors, and are not used while estimating the
orbital parameters. Within each plot, the 57mas radius black disc
represents the resolution limit for $K$-band observations at an 8-metre telescope.}
\label{fig:orbita}
\end{figure*}
\addtocounter{figure}{-1}
\begin{figure*}
\resizebox{0.99\hsize}{!}{{\includegraphics{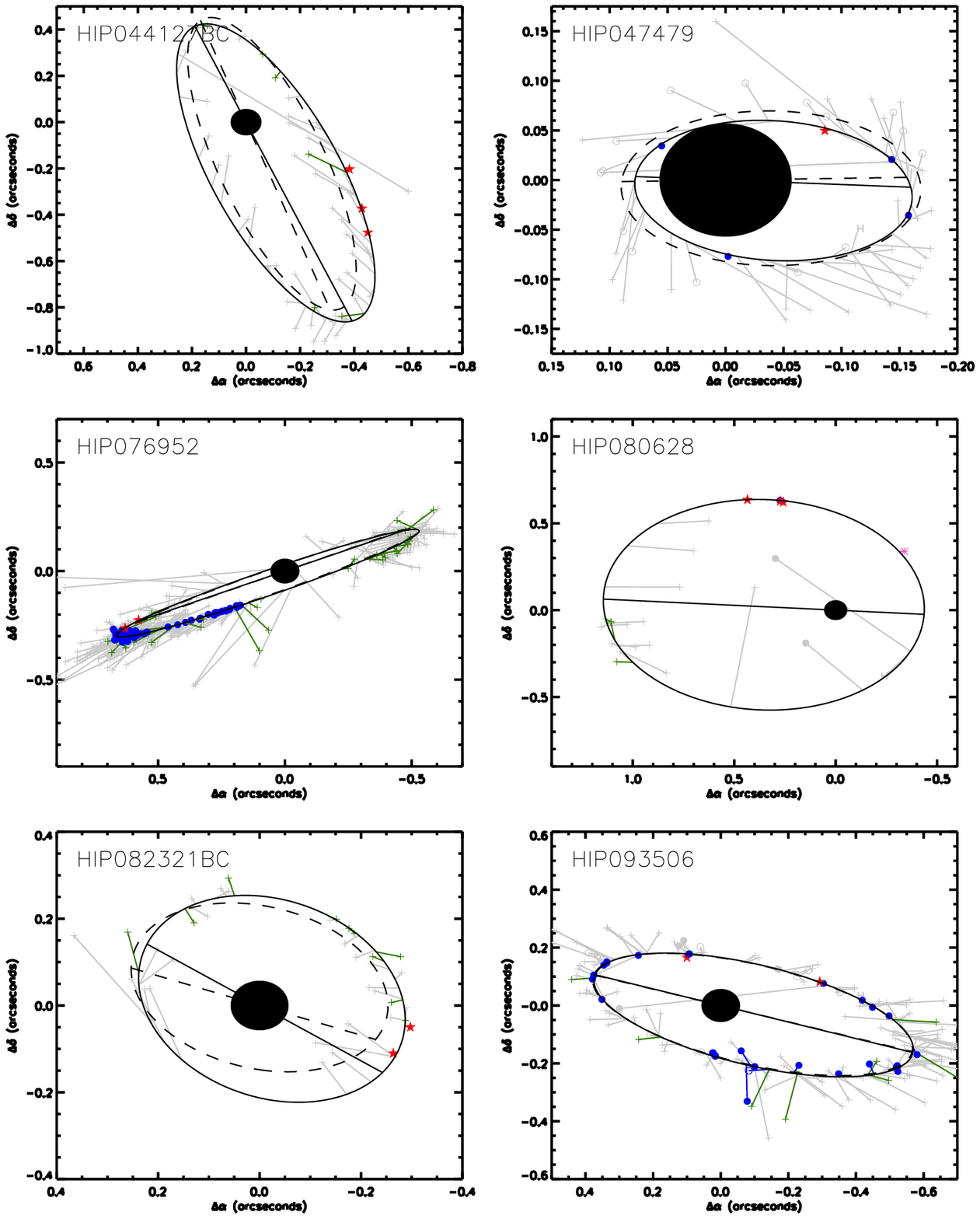}}} 
\caption{Combining our high resolution observations with historical
  measurements, refined orbits for 6 binary systems are plotted. The
  previous orbital fit, obtained from the Sixth Orbit Cataloge, is
  plotted for reference with a dashed line. Each plot uses a similar
  symbol scheme to the Sixth Orbit Catalog; open blue circles representing
  eyepiece interferometry, filled blue circles speckle
  interferometry, green crosses micrometrical observations,
and violet asterisks photographic measurements. Our high resolution
observations presented within this study are plotted as filled red
stars. For each measurement, the corresponding O-C line is plotted,
showing the difference between expected and actual position within
the orbital path. Symbols in grey represent those measurements
presented without formal errors, and are not used while estimating the
orbital parameters. Within each plot, the 57mas radius black disc
represents the resolution limit for $K$-band observations at an 8-metre telescope.}
\label{fig:orbitb}
\end{figure*}
 The orbital parameters for each system with sufficient orbital
 coverage are listed in Table \ref{tab:orbits}, alongside a system
 mass estimated from Kepler's Third Law (hereafter the dynamical
 system mass) and the number of measurements used to fit the orbit. In Figure \ref{fig:orbita}, the refined orbital fits
 incorporating the new data are plotted along with the previously
 reported orbit (references listed within Table
 \ref{tab:orbitsample_obs}). The resulting orbits span a range of
 periods from 10.78 to 467.4 years, and a range of semi-major axes
 between 0$\farcs$12 to 6$\farcs$78. The relatively short period of
 each system allows for orbital motion to be resolved over very short
 baselines, typically on the order of months. The changes to the
 estimated dynamical system mass between the previously published
 orbit fit and the refined fit presented within this study are shown in
 Table \ref{tab:old_new_mass}.

\section{Discussion}
\begin{table*}
\caption{Comparison of dynamical and theoretical system masses for
  each system within the orbit subsample}
\begin{tabular}{lr@{$\pm$}lccr@{$\pm$}lccr@{$\pm$}lccr@{$\pm$}l}
&\multicolumn{2}{c}{Dynamical}&$\cdot$&\multicolumn{3}{c}{\citeauthor{Lejeune:2001fq}}&$\cdot$&\multicolumn{3}{c}{\citeauthor{Marigo:2008fy}}&$\cdot$&\multicolumn{3}{c}{\citeauthor{Siess:2000tk}}\\
&\multicolumn{2}{c}{System}&$\cdot$&Age&\multicolumn{2}{c}{System}&$\cdot$&Age&\multicolumn{2}{c}{System}&$\cdot$&Age&\multicolumn{2}{c}{System}\\
HIP&\multicolumn{2}{c}{Mass (M$_{\odot}$)}&$\cdot$&(Myr)&\multicolumn{2}{c}{Mass (M$_{\odot}$)}&$\cdot$&(Myr)&\multicolumn{2}{c}{Mass (M$_{\odot}$)}&$\cdot$&(Myr)&\multicolumn{2}{c}{Mass (M$_{\odot}$)}\\
\hline
\hline
\multicolumn{15}{c}{Systems identified as doubles (\S 6.1.1)}\\
\hline
5300&3.16&0.34&$\cdot$&280&3.56&0.13&$\cdot$&400&3.47&0.13&$\cdot$&355&3.69&0.16\\
9480&2.72&0.13&$\cdot$&250&3.15&0.20&$\cdot$&280&3.12&0.20&$\cdot$&160&3.25&0.31\\
11569 AaAb$^{\dagger}$&2.12&0.25&$\cdot$&100&2.45&0.07&$\cdot$&100&2.89&0.15&$\cdot$&100&2.97&0.25\\
76952&4.19&0.30&$\cdot$&315&4.38&0.45&$\cdot$&400&4.10&0.40&$\cdot$&315&4.38&0.45\\
\hline
\multicolumn{15}{c}{Low-mass binary pair within hierarchical system
  (\S 6.1.2, \S 6.1.3)}\\
\hline
44127 BC&0.68&0.04&$\cdot$&--&\multicolumn{2}{c}{--}&$\cdot$&250&0.88&0.01&$\cdot$&50&0.83&0.01\\
82321 BC&1.16&0.09&$\cdot$&400&1.91&0.02&$\cdot$&445&1.87&0.02&$\cdot$&500&1.90&0.02\\
\hline
\multicolumn{15}{c}{Systems with one unresolved spectroscopic component
(\S 6.2.1)}\\
\hline
80628&4.99&0.75&$\cdot$&630&3.03&0.08&$\cdot$&630&3.06&0.18&$\cdot$&445&3.15&0.13\\
\hline
\multicolumn{15}{c}{Systems with two unresolved spectroscopic components
(\S 6.2.1)}\\
\hline
28614&6.36&0.62&$\cdot$&630&3.98&0.32&$\cdot$&630&3.96&0.28&$\cdot$&500&4.18&0.34\\
36850&5.42&0.97&$\cdot$&280&4.26&0.28&$\cdot$&355&4.15&0.24&$\cdot$&315&4.34&0.31\\
\hline
\multicolumn{15}{c}{Systems with evidence of unresolved spectroscopic
  components (\S 6.2.2)}\\
\hline
17954&4.15&0.39&$\cdot$&400&3.49&0.11&$\cdot$&400&3.44&0.12&$\cdot$&225&3.67&0.16\\
47479&5.83&0.53&$\cdot$&560&3.99&0.10&$\cdot$&560&3.96&0.10&$\cdot$&445&4.17&0.14\\
93506&5.26&0.37&$\cdot$&710&4.11&0.25&$\cdot$&710&4.07&0.23&$\cdot$&500&4.45&0.36\\
\hline
\multicolumn{15}{l}{$^{\dagger}$ - The ages and theoretical masses may
  be significantly biased due to the presence of additional
  companions}\\
\multicolumn{15}{l}{\;\;\;\; within the resolution limit of the \textsl{Tycho2} and
  \textsl{2MASS} observations.}
\label{tab:comparison}
\end{tabular}
\end{table*}
\begin{figure}
\resizebox{0.99\hsize}{!}{{\includegraphics{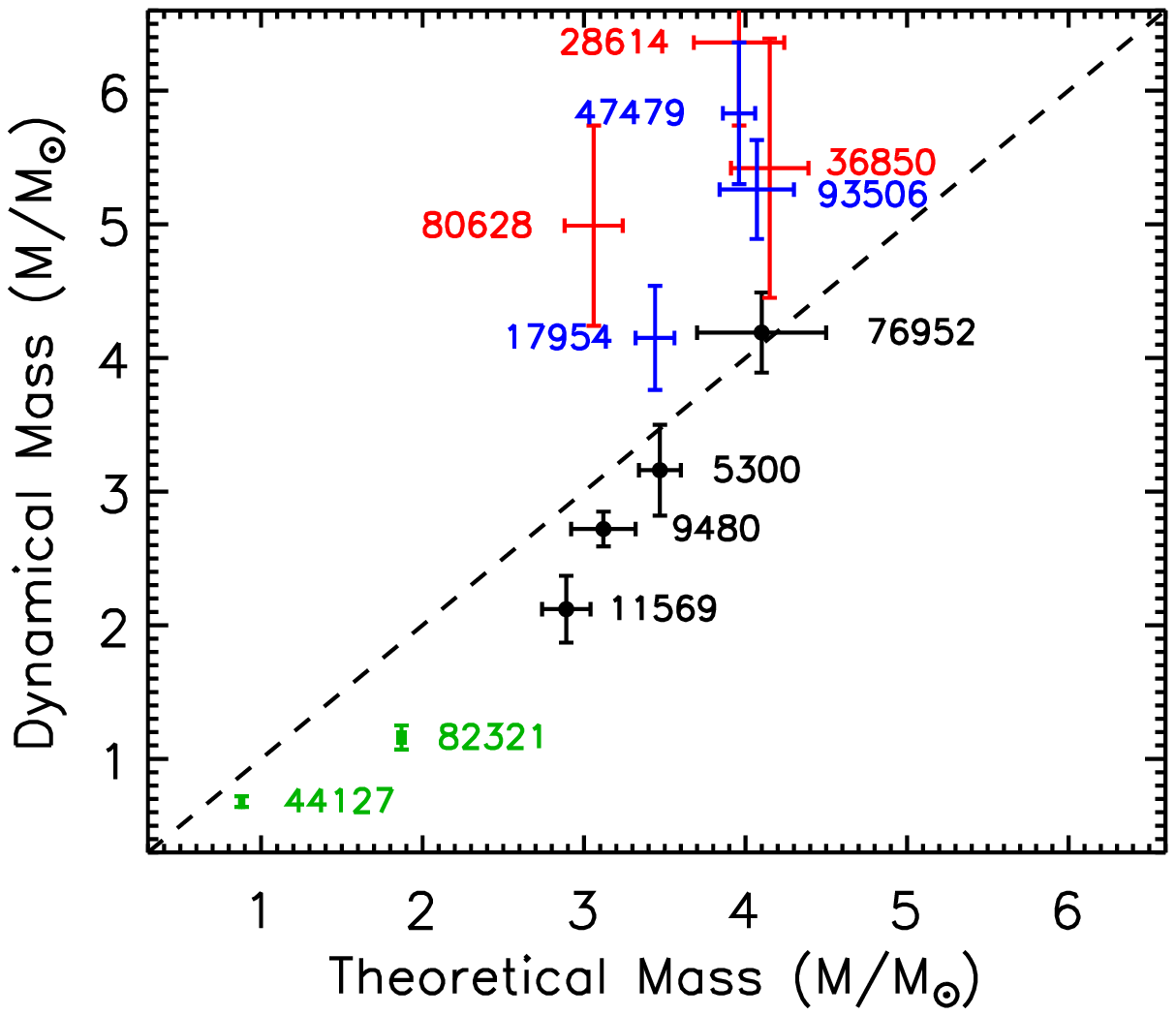}}} 
\caption{Using the method introduced in \S 4.4, a comparison can be
  made between the dynamical mass determined from the orbit and the
  mass estimated from theoretical mass-magnitude relations for the
  stars within the orbit subsample (the \citet{Marigo:2008fy} models
  are used for this example). The systems with an A-type star primary
  which are known to consist only of two components are denoted as
  black points (\S 6.1.1). Two hierarchical systems were fully resolved
  with our data, and the lower-mass pair of each system are in green
  (\S 6.1.2). The systems which have a significantly discrepant
  dynamical mass can be explained by the presence of an unresolved
  spectroscopic companion within our data. The targets with known
  spectroscopic components are plotted in red (\S 6.2.1), while those systems
  with evidence suggesting a previously unknown spectroscopic
  component are plotted in blue (\S 6.2.2). The dashed line denotes the
  equivalence   between the dynamical and theoretical mass.}
\label{fig:comparison}
\end{figure}
We have presented high resolution observations obtained for 26 nearby
multiple systems with A-type primaries with projected separations
within 100 AU. The subset of 12 targets with orbit fits have been
further divided into four distinct categories primarily based on a
comparison between the dynamical mass and the theoretical mass
estimated from the mass-magnitude relations, as shown in Figure
\ref{fig:comparison}. The theoretical mass estimates for each system
using the \citet{Lejeune:2001fq},
\citet{Marigo:2008fy}, and \citet{Siess:2000tk} grids are shown in
Table \ref{tab:comparison}. Those systems with only two known components are
discussed in \S 6.1.1, where the importance of metallicity is
described. Two hierarchical systems resolved within our data are
discussed in \S 6.1.2 and \S 6.1.3, which allow for a comparison to the models
within the K- to M-type spectral range. For systems with a dynamical
mass excess, significantly higher than the mass predicted from the
theoretical mass-magnitude relations, the subset with known
spectroscopic components is discussed in \S 6.2.1, and we present
three systems with evidence suggestive of an additional unresolved
component in \S 6.2.2. The remaining targets are discussed in the
context of continued monitoring of the orbital motion in \S 6.3, of
these 11 are newly resolved as a part of the VAST survey, 2 were resolved in
recent multiplicity surveys of nearby Southern A-type stars
\citep{Ivanov:2006jn,Ehrenreich:2010dc}, and 1 resolved within a large
speckle interferometry survey \citep{McAlister:1987ko}.

\subsection{Comparison to theoretical models}
\subsubsection{A-type binaries}
\begin{figure*}
\resizebox{0.99\hsize}{!}{{\includegraphics{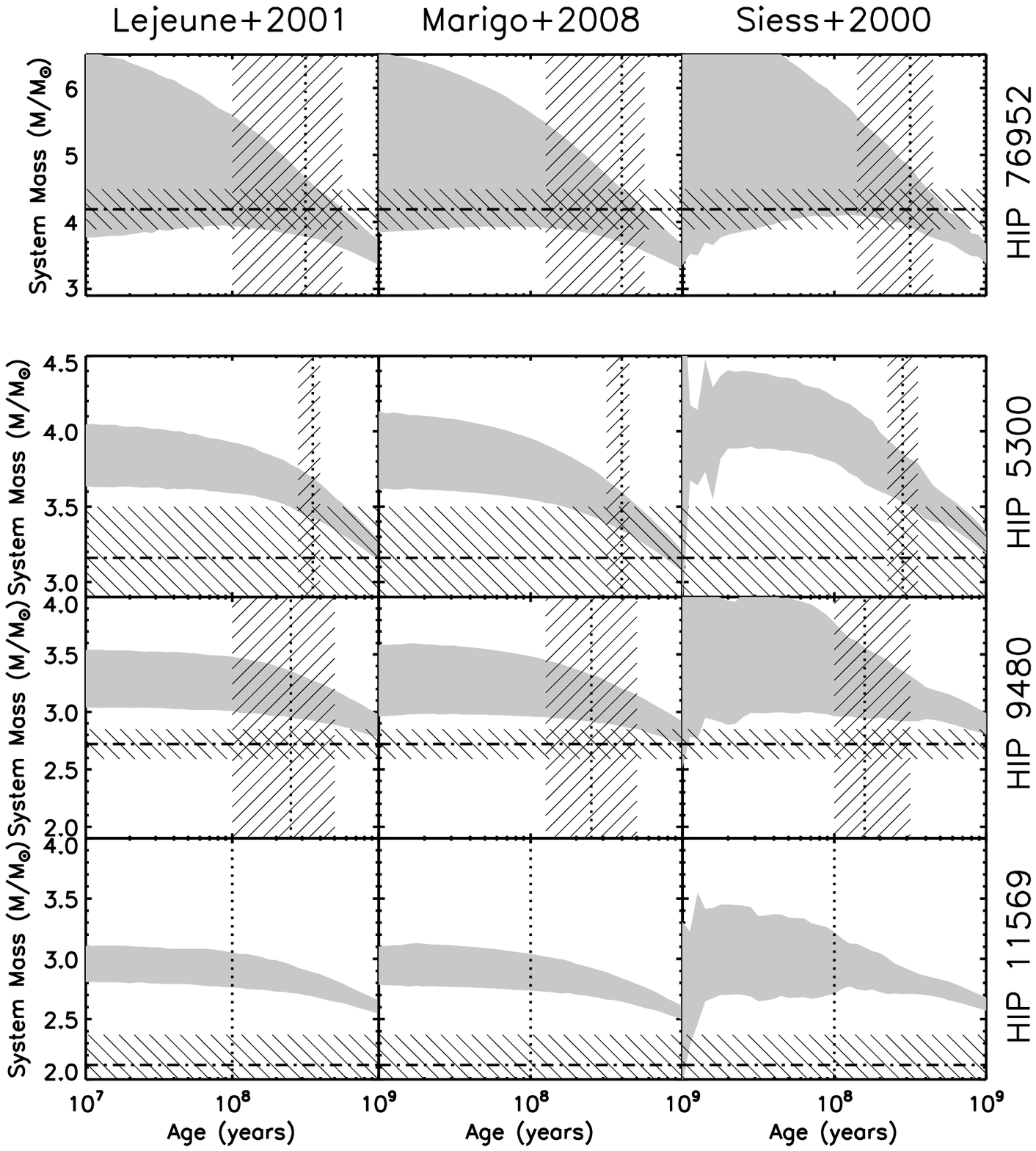}}} 
\caption{The system mass as a function of age based on the
  mass-magnitude relations derived from three of the model grids. Each
panel is similar in nature to Figure \ref{fig:guide}, with the spread
in the mass-age relation introduced by uncertainties in both the
measured $K$-band magnitude, and the distance determination. The
horizontal dot-dashed line indicates the dynamical mass determined
from the system orbit, with the uncertainty denoted by the line-shaded region. The vertical dotted line indicates
the age of the system derived from the isochrones, with the
line-shaded region denoting the range of ages consistent with the
uncertainties in the position of the primary on the CMD. The
  primary of the HIP 11569 system is significantly bluer than expected
  for a Main Sequence star, and as such its age has been assigned to
  100 Myr, with no corresponding uncertainty. The presence
of additional components to the HIP 11569 system, within the resolution
limit of both the \textsl{Tycho2} and \textsl{2MASS} observations, is
the likely cause of this bias.}
\label{fig:true}
\end{figure*}
\begin{figure}
\resizebox{0.99\hsize}{!}{{\includegraphics{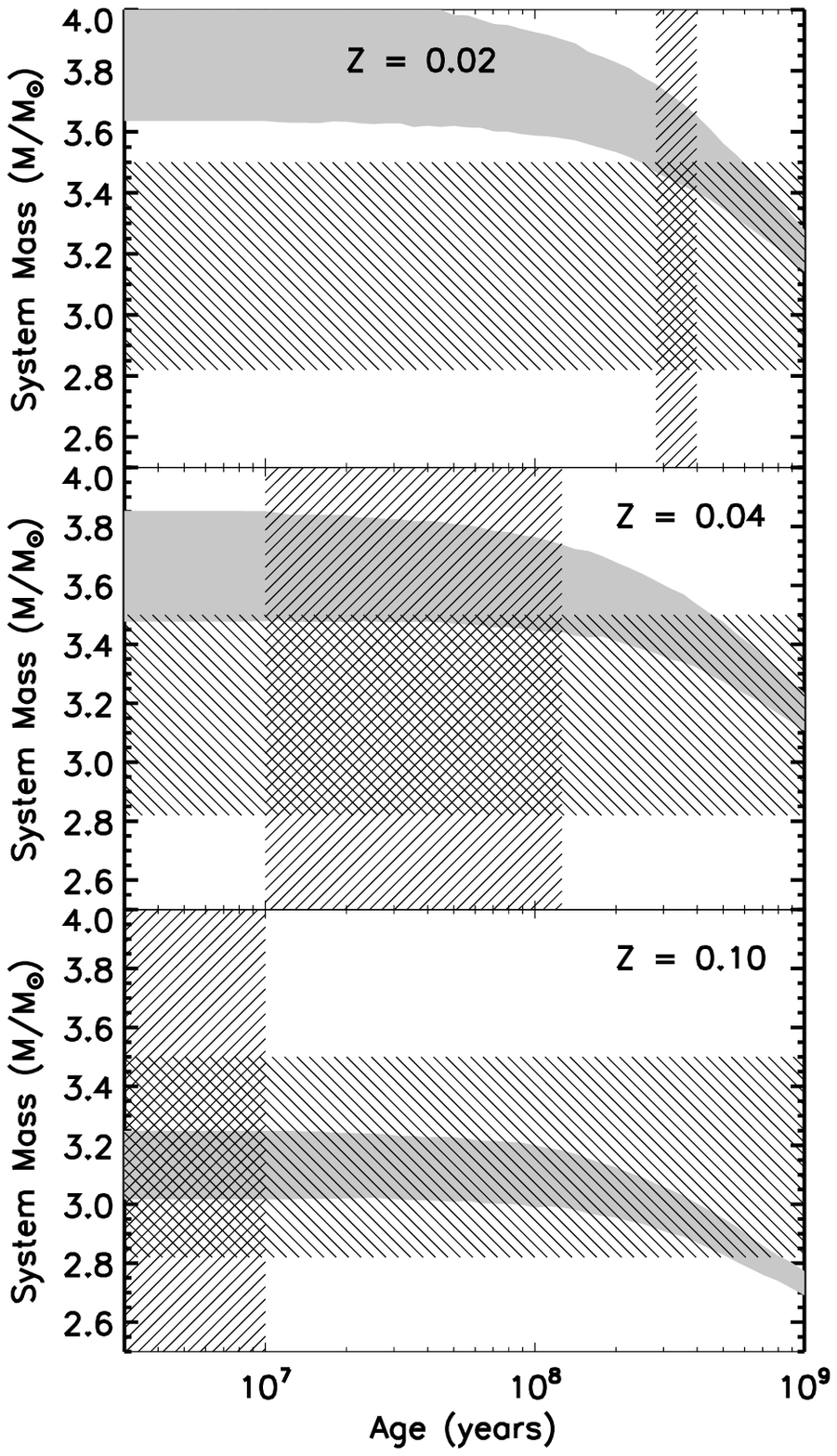}}} 
\caption{\textit{(top panel)}: The dynamical mass of the HIP 5300
  system (horizontal line-shaded region) is significantly below that of the system mass estimated from
  mass-magnitude relations (solid shaded region) calculated from the Solar metallicity
  models from \citet{Lejeune:2001fq}. \textit{(middle panel)}: Assuming
a higher metallicity ($Z=0.04$), the position of the components on the
colour-magnitude diagram suggests a younger age (denoted by the
vertical line-shaded region), and the
mass-magnitude relation derived from the metal-enhanced models
suggests a systematically lower system mass, although the dynamical
mass is still discrepant. \textit{(bottom panel)}: With a metallicity
of $Z=0.10$, the age estimate is younger still ($<$10 Myr), and the
discrepancy between the dynamical mass and the mass estimated from the
mass-magnitude relation is removed.}
\label{fig:metallicity}
\end{figure}
Four targets within the orbit subsample are systems where the two
known components have been resolved within our high resolution data; HIPs
5300, 9480, 11569, and 76952. For each of these four targets, the
mass-magnitude relations were used to determine how the system mass changes as a function of
our estimate of the system age, shown in Figure \ref{fig:true}. For
one system, HIP 76952, the dynamical system mass is consistent with
the theoretical system mass (Figure \ref{fig:true}, top panel). The
ages of the systems estimated from the position of the primary on the
CMD are consistent with their position within the Local Interstellar
Bubble (LIB). The minimum age of a star within the LIB, excluding
those with relatively high space motions (e.g. the $\beta$ Pic moving
group - \citealp{Ortega:2002je}), has been shown to be 160 Myrs
\citep{Abt:2011ha}.

The dynamical system masses of the three remaining systems are consistently
lower than their theoretical system masses. One possible explanation
for the apparently low dynamical masses is a non-Solar
metallicity. Varying the metallicity has a 
significant effect on the system age estimate and the mass-magnitude
relations. As an example, a 2 M$_{\odot}$ star with super-Solar
metallicity will be more luminous and have a redder $V-K$ colour index
than a similar-mass star of Solar metallicity. A super-Solar metallicity star will appear to be significantly older
based on its position on a CMD if Solar metallicity models are
used. To explore the effect of our assumption of Solar metallicity for
the entire sample, the HIP 5300 system (Figure \ref{fig:true}, second
panel) was studied at varying metallicity values. Using the Solar
metallicity models, the dynamical system mass of this system is
significantly lower than the theoretical system mass
(Figure \ref{fig:metallicity}, top panel). Increasing the
metallicity causes the star to appear both younger, and
less massive (Figure \ref{fig:metallicity}, bottom panel). Only eight of
the targets included within this study have metallicity measurements,
either from spectroscopic analysis (e.g. \citealp{Erspamer:2003hz}) or
derived from Str\"{o}mgren photometry (e.g. \citealp{Song:2001bv}),
demonstrating the need for further study in this area.
\subsubsection{K- and M-type binaries - HIP 44127}
\begin{figure*}
\resizebox{0.99\hsize}{!}{{\includegraphics{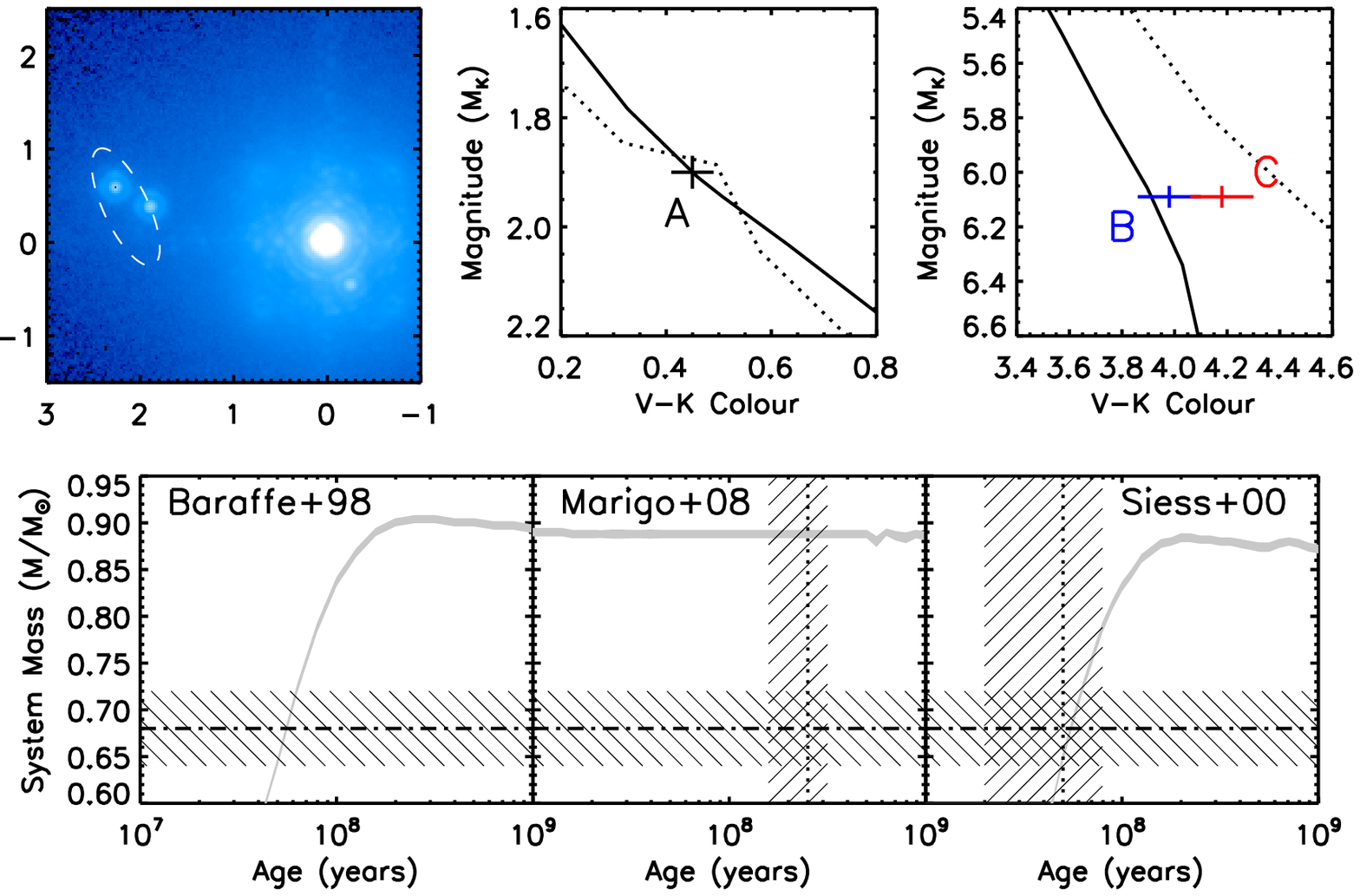}}} 
\caption{\textit{Top row} - \textit{(left)}: An AO observation of HIP
  44127 triple system. The orbit of the lower-mass pair (BC) taken from
Table \ref{tab:orbits} is plotted as a solid line for reference. The
bright point source to the lower-right of the primary PSF is a known
filter ghost. The axis labels are given in arcseconds, with each
arcsecond equal to approximately 14.5 AU. \textit{(centre)}: A colour-magnitude diagram highlighting the
position of the primary with respect to the theoretical models;
\citet{Marigo:2008fy} - solid line, \citet{Siess:2000tk} - dotted
line, while the \citet{Baraffe:1998ux} models do not extend to high
enough masses. The age of the system was estimated using the isochrone
which best fit the position of the primary. \textit{(right)}: A colour-magnitude diagram
highlighting the position of the two lower-mass components (B - blue,
C - red). The line styles are the same as the centre panel \textit{Bottom
  row} - The theoretical system mass plotted as a function of system
age for each of the three models. The vertical dotted line within each panel
indicates the age of the system estimated from using each of the
models, while the horizontal dot-dashed line indicates the dynamical
system mass estimated from the orbital fit. As the
\citet{Baraffe:1998ux} models do not extend to the A-type star mass
range, an age estimate is not possible.}
\label{fig7}
\end{figure*}

The detection of two hierarchical systems, with pairs of lower-mass
companions in a wide orbit around an A-type primary, allows for a
comparison of the theoretical models in the low-mass regime where the
models differ in the treatment of the contraction phase of these
objects (e.g. Figure \ref{fig:massmag}). Our observation of the HIP
44127 system, shown in Figure \ref{fig7}, resolves three components to this hierarchical system,
with an A-star primary (A) separated by $\sim 4\arcsec$ from two
fainter, gravitationally bound, companions (BC).

The orbit presented in the Sixth Orbit Catalog of the BC pair around
the A-type primary is in disagreement with our recent observations of
this system. Although the phase coverage is
insufficient for a robust orbital determination, the high proper
motion of the primary ($\Delta \alpha = -441.1$ mas yr$^{-1}$, $\Delta
\delta = -215.2$ mas yr$^{-1}$) suggests that the BC pair is
co-moving. In addition, radial velocity variations detected within the spectra of the primary
suggest the presence of a spectroscopic component to this system with
a period of $\sim$11 years \citep{Abt:1965fz}. The separation of this
component was anticipated to be between 0$\farcs$2 and 0$\farcs$6,
reaching maximum separation in the middle of 2007
\citep{Docobo:2006bh}. We do not resolve any companion in any epoch
of our observations which are consistent with these predictions,
placing an upper limit to the separation of 0$\farcs$08, 0$\farcs$10,
and 0$\farcs$10 in 2008, 2010, and 2011 respectively. The magnitude
difference between the primary and the suggested companion,
estimated to be $\Delta m=1.2$, is well within our detection limits at
the expected separation \citep{Docobo:2006bh}.

A refined orbit of HIP 44127 BC is presented in Table \ref{tab:orbits} and Figure \ref{fig:orbitb},
with an estimated dynamical system mass of 0.68$\pm$0.04
M$_{\odot}$. The small magnitude difference between the two components
($\Delta K \sim 0$) suggests a mass ratio close to unity. Assuming
individual masses of $\sim 0.34$ M$_{\odot}$ the components would be of
early- to mid-M spectral type \citep{Baraffe:1996ip}, a region of particular
disagreement between theoretical models. The
\citet{Baraffe:1998ux} and \citet{Siess:2000tk} models both predict
that for a binary consisting of two stars of magnitudes equal to the
measured magnitudes of the two components ($M_K=5.87$), the
system mass will increase from 0.3 to 0.9 M$_{\odot}$ between 10 and
100 Myrs (Figure \ref{fig7}), as these models take into account the
phase of contraction onto the Main Sequence for lower-mass stars. No
such change is predicted by the \citet{Marigo:2008fy} models, with the
system mass remaining unchanged between 10 Myrs and 1 Gyr
(Figure \ref{fig7}). The theoretical system masses are only consistent
with the dynamical mass at $\sim 30$ Myrs in the
\citet{Baraffe:1998ux} and \citet{Siess:2000tk} models, and are not
consistent with any age in the \citet{Marigo:2008fy} models. The
metallicity of this system has been measured to be almost Solar
\citep{Wu:2010cu}, removing the degeneracy which exists between
metallicity and the mass and age of the system (e.g. \S 6.2.6). 

The position of the primary on the CMD suggests a
relatively young age of the system between 50 and 250 Myr. The $V-K$
colours of the BC pair also suggest a young age, between 40 and 100 Myr
using the \citet{Siess:2000tk} and \citet{Baraffe:1998ux} models. The
inferred young age of the system is not inconsistent with the minimum
age of stars found within the LIB \citep{Abt:2011ha}, given the
relatively high $UVW$ space velocity of the HIP 44127 system
\citep{Palous:1983ty}. The detection of X-ray emission from this
system is also of interest. Previous studies have shown that A7 stars
such as the primary should not emit X-rays, and that any detection of X-rays from
the position of the star can be indicative of a lower-mass companion
(e.g. \citealp{DeRosa:2011ci}). The \textsl{ROSAT} source
J085913.0+480227 is coincident with the optical position of HIP 44127
\citep{Voges:1999ws}.

This hierarchical triple system warrants further study, primarily in
order to refine the orbital fit as the BC pair approaches apastron
passage in 2018. This system also makes an ideal candidate for future
spectroscopic observations to search for the
narrow spectral lines of the two faint companions. With a double-lined
spectroscopic orbit fit, model-independent masses can be calculated
for the individual components. If the young age suggested by both the
position on the CMD and the system mass estimated
from the mass-magnitude relations is correct, the BC pair would be an
ideal calibrator for the theoretical models in the young, low-mass
regime.

\subsubsection{K- and M-type binaries - HIP 82321}
\begin{figure*}
\resizebox{0.99\hsize}{!}{{\includegraphics{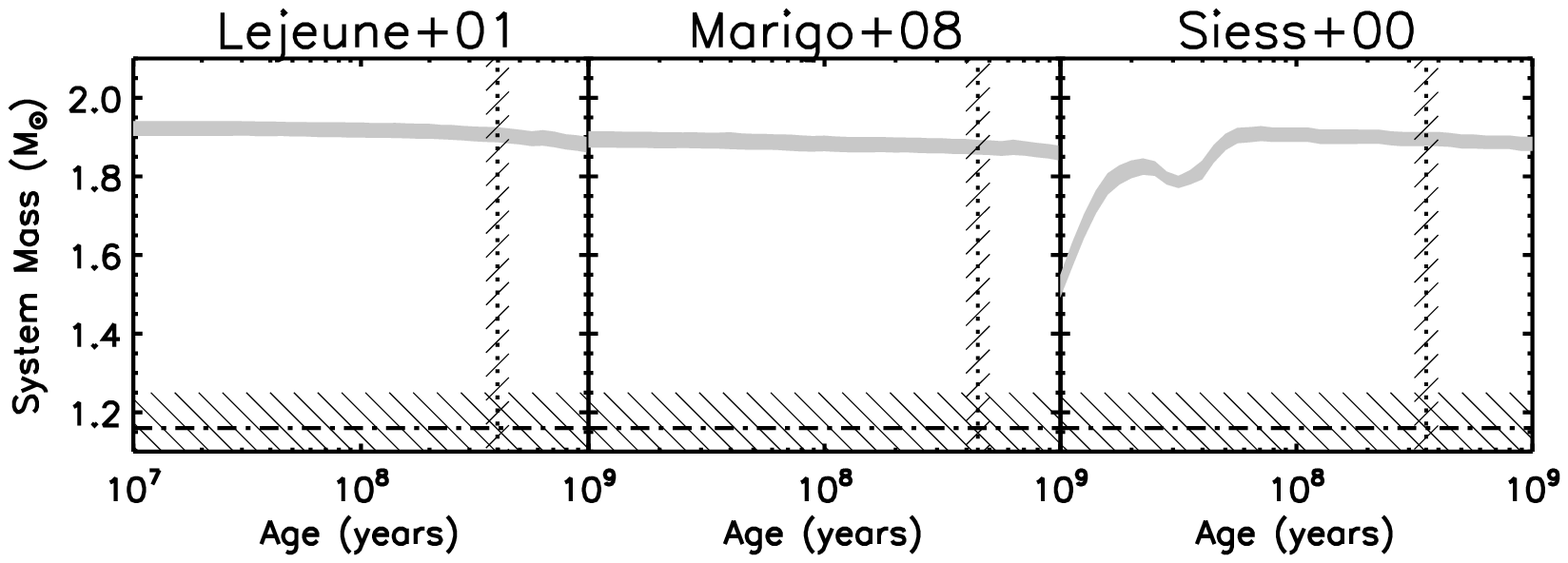}}} 
\caption{The theoretical system mass plotted as a function of system
age for each of the three models for the HIP 82321 BC system. The
vertical dotted line within each panel indicates the age of the system
estimated from using each of the models, while the horizontal
dot-dashed line indicates the dynamical system mass estimated from the
orbital fit. The clear discrepancy between the dynamical and
theoretical system masses can be explained by the lack of complete
coverage of the orbit presented in Figure \ref{fig:orbitb}. Further
high angular resolution observations of this system will allow for an
improved determination of the orbital parameters.}
\label{fig:82321}
\end{figure*}

The hierarchical triple system HIP 82321 was resolved within a single
epoch of our AO observations. The A-type primary (A) is $\sim
2\arcsec$ from a binary pair of two lower-mass companions (BC) in
a wide orbit. The significant proper motion of the primary ($\Delta  \alpha
= 22.8$ mas yr$^{-1}$, $\Delta \delta = -51.4$ mas yr$^{-1}$), and near
constant separation of the three stars, suggest that the lower-mass pair
is co-moving with the primary. Spectroscopic measurements of the
lower-mass components of this system are possible given the small
magnitude difference of the pair with respect to the primary ($\Delta
K \sim 2.5$); such spectra would allow for a determination of the
individual masses independent of the distance to the system. The
\textsl{Hipparcos} parallax and \textsl{2MASS} $K$-band 
magnitude measurements of this system allow for a tight constraint of
the system age to between 300 and 400Myrs, based on the position of
the primary on the CMD. The primary is also a possible member of the Ursa Major
moving group, an association of Solar metallicity stars with an age
between 300 Myrs \citep{Soderblom:1993kr} and 500 Myrs
\citep{King:2003df}.

A refined orbit of the BC pair is
presented in Table \ref{tab:orbits}, with an estimated dynamical
system mass of 1.32$\pm$0.05 M$_{\odot}$. The theoretical system mass,
assuming the distance to each component is the same as the primary, is
significantly higher at $\sim 2.0$ M$_{\odot}$ (Figure \ref{fig:82321}). The sparse coverage of the orbit
(Figure \ref{fig:orbita}) suggests that the orbit fit could be poor,
resulting in an incorrect dynamical system mass. The orbit fit would be
significantly improved with subsequent observations. An alternative
scenario is that the pair are a background binary with a proper
motion similar to the primary, which would bring the dynamical
and photometrically derived masses into agreement. 

\subsection{Higher-order multiplicity}
\begin{figure}
\resizebox{0.99\hsize}{!}{{\includegraphics{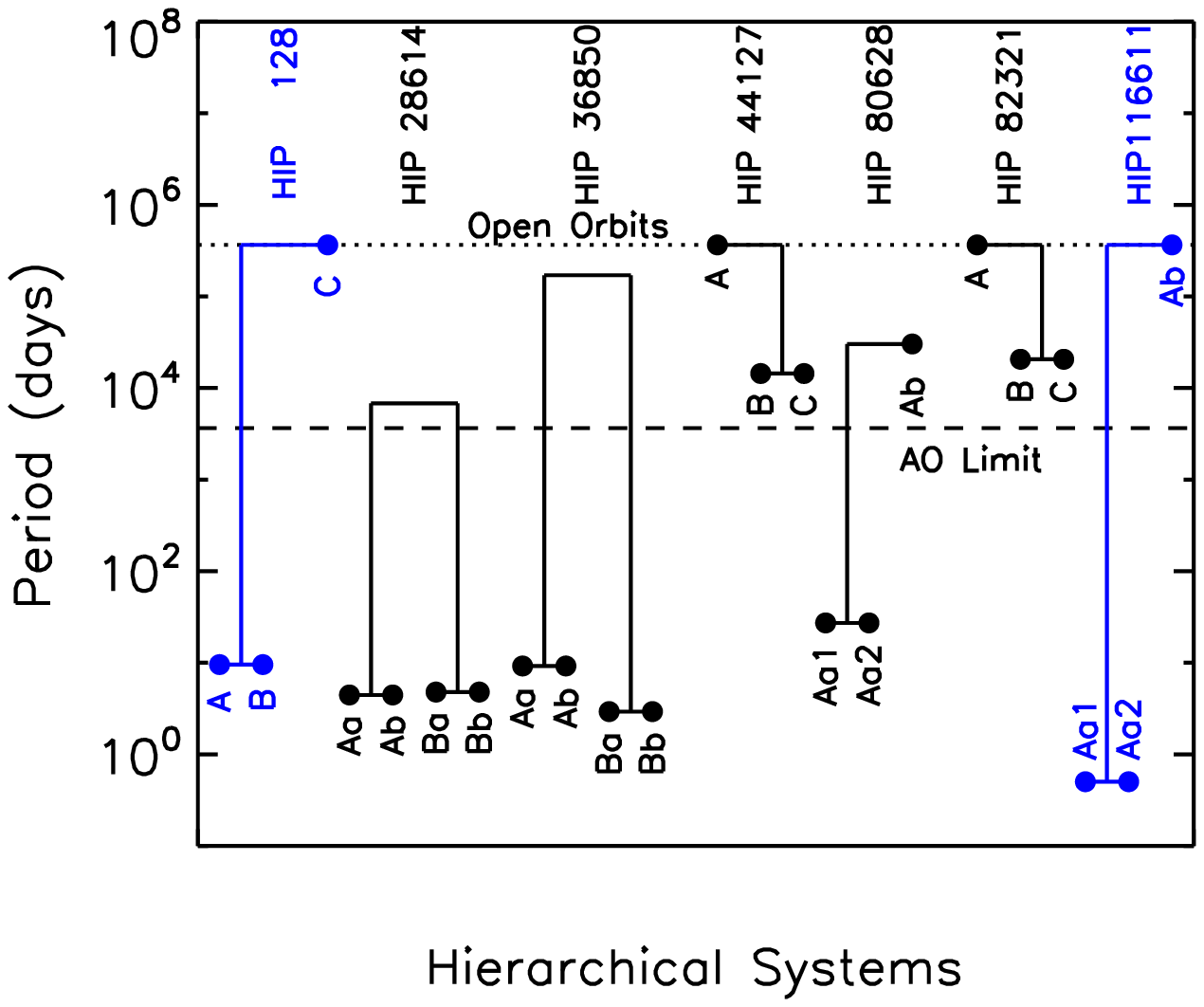}}} 
\caption{A schematic of the 7 hierarchical systems within both
  subsamples, including only components within a projected separation
  of 100AU to the A-type primary. The schematic diagram does not
  include the suspected spectroscopic binaries described in \S 6.2.2. Those systems which were newly
  resolved as a part of the VAST survey are highlighted in blue. Each filled circle represents an
  individual component, with the period of the binary pair denoted by
  its vertical position. Our high resolution observations
are only sensitive to systems with orbital periods greater than $\sim$
10 years, denoted by the dashed horizontal line. For wide
separation systems with indeterminate orbital periods, such as HIP
128 AB--C, the orbital period has been arbitrarily set at 1000 years,
denoted by the dotted horizontal line.}
\label{fig:hierarchical_line}
\end{figure}
\subsubsection{Known spectroscopic binaries}
\begin{figure*}
\resizebox{0.99\hsize}{!}{{\includegraphics{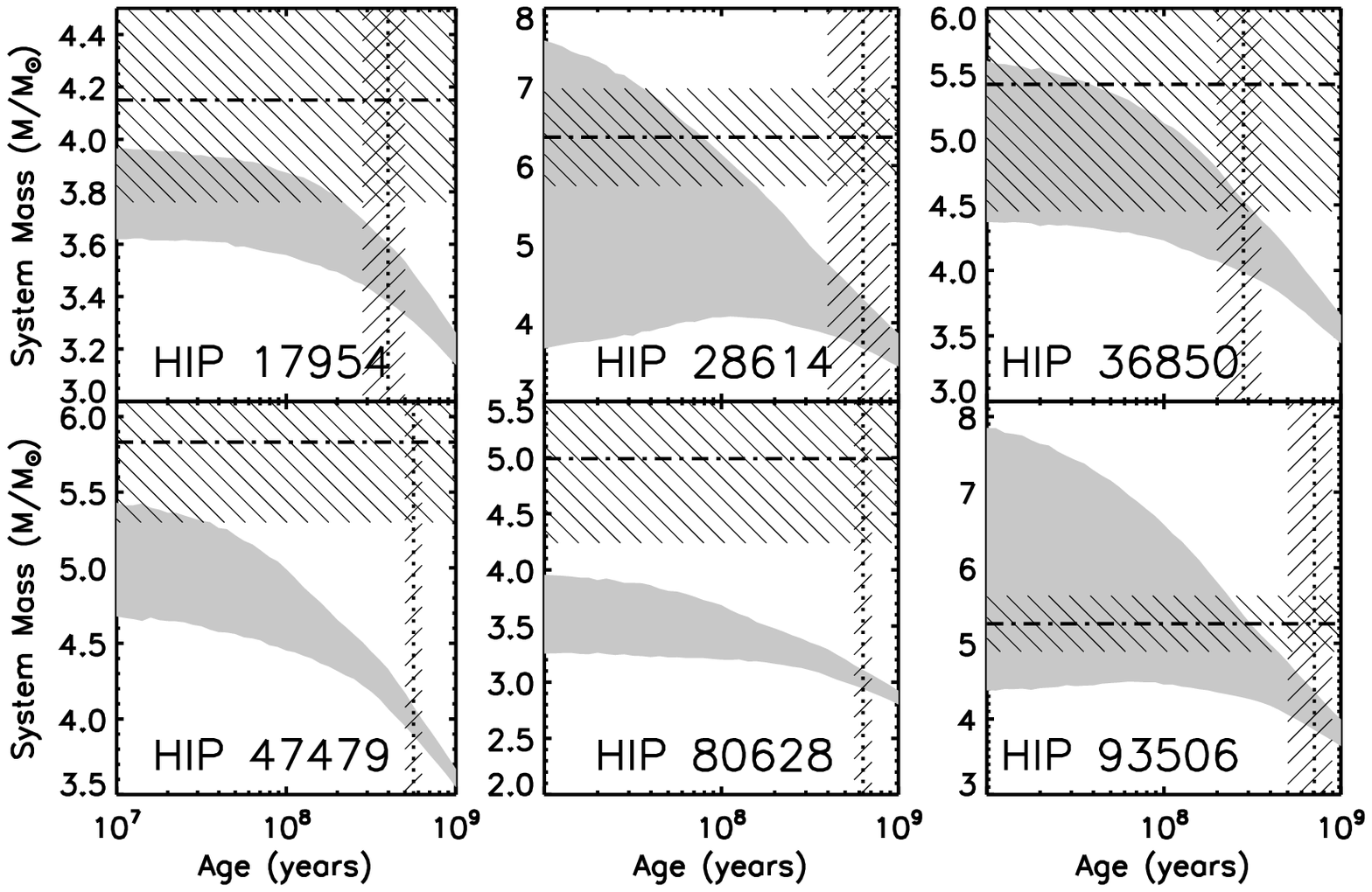}}} 
\caption{The six systems for which spectroscopic components are either known
  to be present through previous spectroscopic observations (HIPs
  28614, 36850, 80628), or thought to exist due to a significant
  discrepancy between the theoretical system mass and the dynamical
  system mass. For clarity only one model has been used to estimate
  the theoretical system mass \citep{Lejeune:2001fq}. The horizontal
  displacement between the intersection of the dynamical system mass
  (dot-dashed line) and the age estimate of the system (dotted line), and
  the theoretical system mass curve (solid shaded region), gives an
  order of magnitude estimate of the mass of the unresolved
  spectroscopic components.}
\label{fig:spectroscopic}
\end{figure*}
The presence of an unresolved spectroscopic binary can have a
significant impact on the magnitude assigned to each component of the
multiple system, increasing the estimated component magnitude by as
much as 0.75 mag for an unresolved equal-mass spectroscopic
binary. For each target within this study, the literature was searched for
references to additional components resolved through spectroscopic or
interferometric observation which would influence any comparison made
between the dynamical system mass and the theoretical system mass
(e.g. Figure \ref{fig:spectroscopic}). Of
the targets within the orbit subsample, three are known to have
additional spectroscopic components (HIPs 28614, 36850,
80628). The spectroscopic component to HIP 9480 resolved by
  \cite{Abt:1965fz} is resolved within our adaptive
  optics observations. Similarly, for the monitoring subsample, two are known
spectroscopic binaries (HIPs 128, 116611). The greater frequency of
  spectroscopic binaries within the orbit sample, relative to the
  monitoring sample, can be explained by the narrow spectral lines
  of the former sample due to their relatively low radial velocities,
  $\langle v\sin i\rangle = 71$ km s$^{-1}$ compared with $\langle v\sin
  i\rangle = 115$ km s$^{-1}$ for the latter sample
  \citep{Abt:1995eo}. The magnitudes listed for
the resolved components of these systems (Table \ref{tab:components})
are the blended magnitudes of the listed spectroscopic components. In
addition to these known spectroscopic binaries, three members of the
orbit subsample are resolved as hierarchical triples within our
high resolution observations (HIPs 11569, 44128, 82321). A schematic
representation of the higher-order multiplicity systems is given in Figure
\ref{fig:hierarchical_line}.

For the three systems with known spectroscopic components within the
orbit sample, the dynamical system mass includes the mass of each
component, regardless of whether it is resolved within our data. This
causes a significant discrepancy when the dynamical system mass is compared with the
theoretical system mass (e.g. Figure \ref{fig:comparison}), with the
dynamical system mass being systematically higher. The discrepancy
between the two values cannot be directly converted into a mass for
the unresolved components however, as a blended magnitude would have been
used when determining the mass from the mass-magnitude relations.

\subsubsection{Evidence of spectroscopic components}

Systems with significantly higher dynamical
masses than theoretical masses obtained from mass-magnitude relations
are strong candidates for multiple systems with unresolved
components which have not been detected in spectroscopic
observations. We find two such systems with the
signature of a possible unresolved component: HIP 17954 and HIP
93506 (Figure \ref{fig:spectroscopic}), with masses of the order of 0.5 and 1.5
M$_{\odot}$. Sensitive spectroscopic observations of both systems may
lead to the detection of the spectral lines from an unresolved
lower-mass component. A similar phenomenon is observed for the HIP
47479 system, although the number of measurements used to
determine the orbit is particularly low, making the dynamical mass
less certain. The narrow spectral lines of HIP 47479, implied by the
low measured stellar rotational velocity \citep{Royer:2007dj}, make it
an ideal candidate for spectroscopic follow up. Previous spectroscopic
observations of this system reveal $v\sin i$ variations with a
magnitude of 40 km s$^{-1}$ \citep{Moore:1932vo}.

From the total sample of 26 systems, and only considering stellar
companions within 100 AU, a lower-limit on the higher-order
multiplicity of A-type stars can be estimated. Assuming the suspected
unresolved companions described earlier in this section are true,
there are five double, six triple, and two quadruple systems within
the orbit subsample, corresponding to frequencies of 39, 46, and 15
per cent, respectively. This lower-limit shows an enhancement on the
higher-order multiplicity of A-type stars when compared with
Solar-type primaries (74 per cent double, 20 per cent triple, and 6
per cent quadruple or higher-order - \citealp{Raghavan:2010gd}), and
is more consistent with the fraction reported for more massive O-type
primaries (46 per cent double, and 54 per cent triple or higher-order
- \citealp{Mason:2009fm}).

\subsection{Continued monitoring targets}
\subsubsection{Newly-resolved binaries}
\begin{figure*}
\resizebox{0.9\hsize}{!}{{\includegraphics{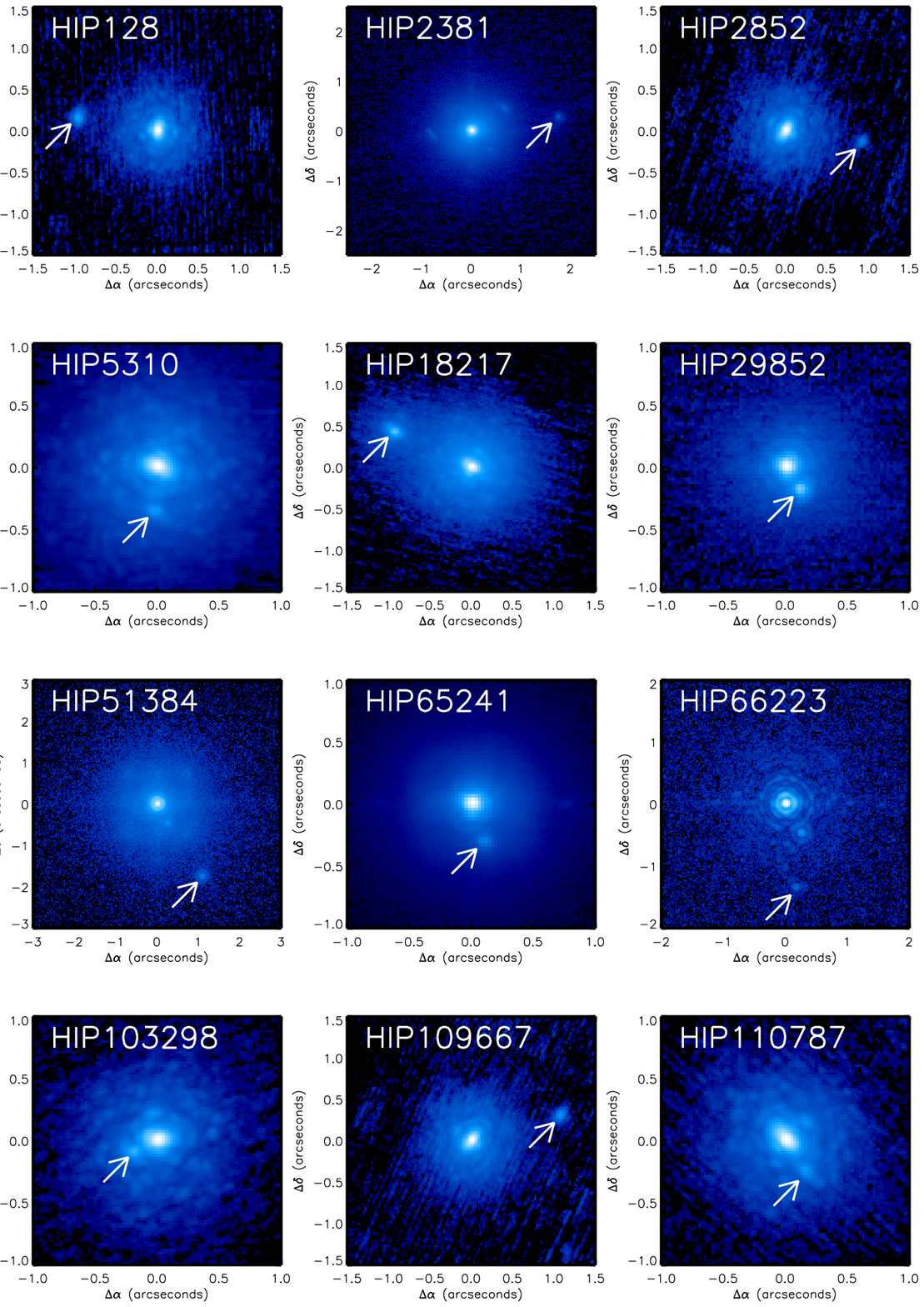}}} 
\caption{Observations of twelve of the thirteen systems with projected
  separations $< 100$AU suggested as future orbital monitoring
  targets. The companion in each image is highlighted, with a
  logarithmic image scale between $1$ (white) and $10^{-6}$ (black)
  relative to the peak intensity of the primary.}
\label{fig:futurea}
\end{figure*}
\addtocounter{figure}{-1}
\begin{figure*}
\resizebox{0.9\hsize}{!}{{\includegraphics[trim=0mm 152mm 0mm 0mm]{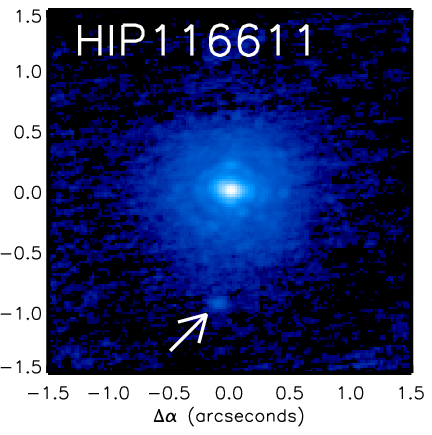}}} 
\caption{\textit{(continued)}: An observation of one of the thirteen systems with projected
  separations $< 100$AU suggested as future orbital monitoring
  targets. The companion in each image is highlighted, with a
  logarithmic image scale between $1$ (white) and $10^{-6}$ (black)
  relative to the peak intensity of the primary.}
\label{fig:futureb}
\end{figure*}
We have identified 13 binary systems with projected separations
ranging between 13 AU and 96 AU which would make ideal candidates
for future orbital monitoring projects (Table
\ref{tab:futuresample_obs}). The 100 AU projected separation cut-off
was applied to select only systems for which orbital motion could be
detected with several years of observations. The binaries resolved
within the monitoring subsample typically have lower-mass ratios than for the
orbit subsample, a demonstration of the effectiveness of AO
observations at detecting high-contrast binaries. Based on their
position on the CMD, two  members of the
monitoring subsample (HIP 5310, HIP 18217) appear to lie on the Zero-Age
Main Sequence, and the measured magnitude difference between primary
and secondary would correspond to a late K or early M-type companion in each
case. These companions are of particular interest as they will allow
for further tests of the theoretical models within the young, low-mass
regime. A gallery of the observations obtained of each target within
the monitoring subsample is shown in Figure\ref{fig:futurea}.

\subsubsection{HIP 77660}
\begin{figure}
\resizebox{0.99\hsize}{!}{{\includegraphics{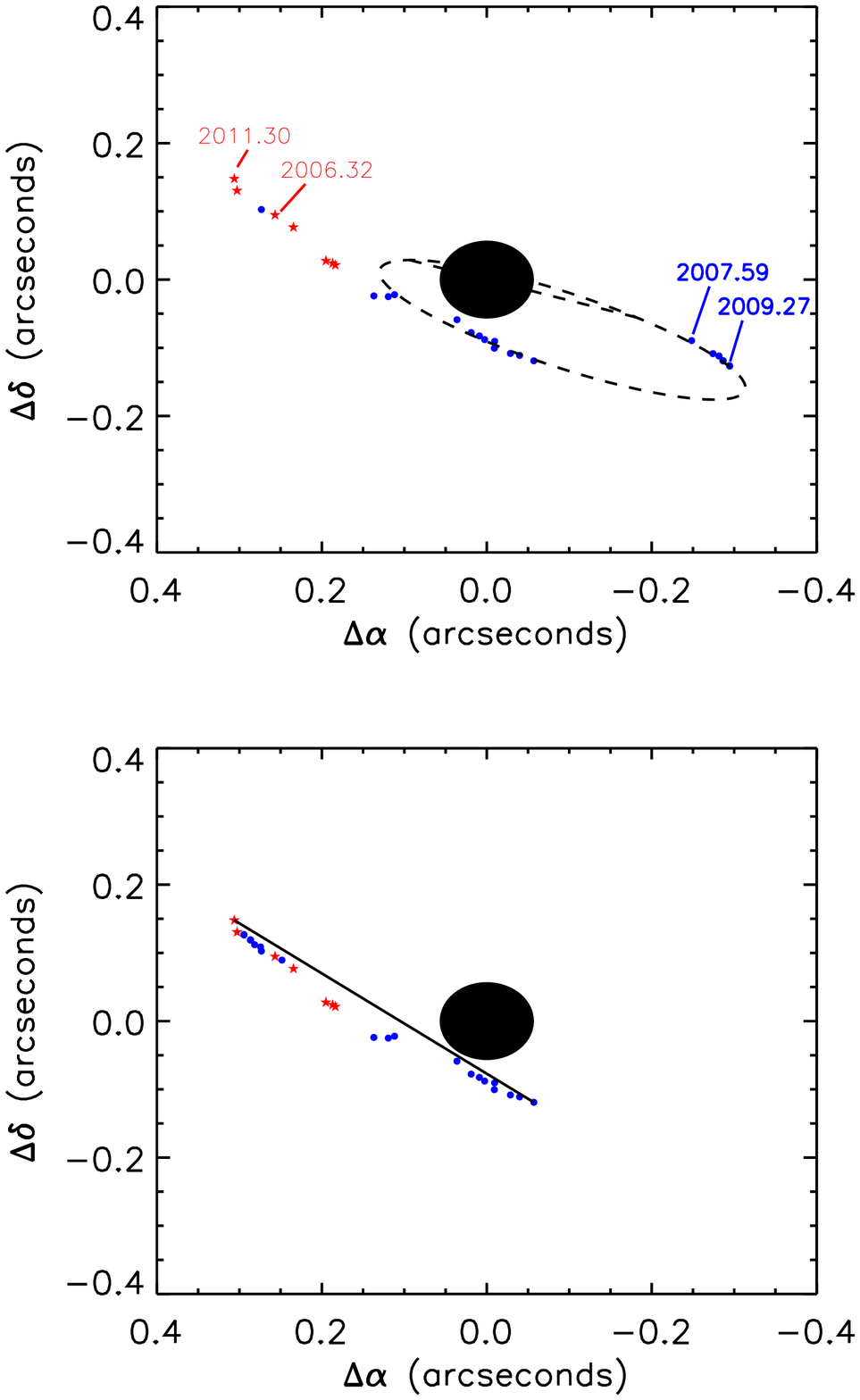}}} 
\caption{\textit{(top panel)}: The historical measurement of the
  system HIP 77660, plotted
  with the orbital fit estimated by \citet{Docobo:2010hf}. This
  system demonstrates the quadrant ambiguity limitation of the
  speckle interferometry technique. The binary companion was measured
  by \citet{Tokovinin:2010ft} using speckle interferometry to be at a
  position angle of 248.4$^{\circ}$ to 246.8$^{\circ}$ between
  mid-2007 and mid-2009 (blue filled circles, top right). Our AO
  images, which do not suffer from the same quadrant ambiguity, shows
  the binary at a position angle of 69.7$^{\circ}$ to 67.3$^{\circ}$ between
  mid-2006 and mid-2011 (red filled stars, left). \textit{(bottom
    panel)}: The relative position of the secondary, with 180$^{\circ}$ added to the four speckle
  interferometry measurements which were inconsistent with the
  observations presented within this study. The solid line connects the first and
  last observations, showing a deviation away from linear
  motion. Further observations are required to determine if this
  deviation is indicative of the true orbital path of the secondary
  about the primary.}
\label{fig8}
\end{figure}
An advantage of monitoring binary systems for orbital motion with AO
instrumentation over interferometric techniques is the elimination of the
quadrant ambiguity. In some cases, the output of the image processing
of speckle data results in a 180$^{\circ}$ ambiguity on the position
angle measurement \citep{Bagnuolo:1992ck}. This may lead to a
scenario where orbital motion is thought to exist for a binary pair,
when in fact this motion is the product of such a quadrant uncertainty
combined with the true linear motion of the companion. The ambiguity
can be resolved by observing the system using AO imaging, where no
reconstruction is required to obtain the final science image.

Based on our AO observations of the binary system HIP 77660, it appears
that such a quadrant ambiguity has occurred, making linear motion
appear as orbital motion, as shown in Figure \ref{fig8}. The
significant proper motion of the system ($\Delta\alpha = -91$ mas/yr,
$\Delta\delta = -28$ mas/yr), as measured by \textsl{Hipparcos}, is
inconsistent with a stationary background object. While future
measurements of this system will be able to resolve the presence of
orbital motion after a sufficient time baseline, spectroscopy or multi-colour
photometry will allow for a rapid characterisation of the properties
of both components.

\section{Summary}
We have presented high resolution observations of 26 nearby multiple
systems with A-type primaries with projected separations $<$ 100 AU,
11 of which are binaries newly resolved as a part of the VAST survey. For those
systems with sufficient orbital motion coverage, refined orbital
parameters were calculated and the estimated dynamical system mass was
compared to masses derived from theoretical models. Due to their rapid
evolution across the CMD, removing the significant age degeneracy for
lower-mass solar-type stars, binaries with A-type components are ideal
targets with which to test theoretical models. Four such systems were
investigated, with one system having consistent dynamical and
theoretical system mass estimates. Of the remaining three systems,
each had a dynamical mass significantly lower than that predicted from
the models. While this discrepancy may be indicative of a true
divergence between the models and observations, the lack of
metallicity measurements for these systems provide another
explanation. Future orbital monitoring observations of A-star binary 
systems will provide further refinement to the orbital parameters and,
combined with refinement of the magnitude, metallicity, and parallax
measurements, will improve the analysis performed within this study.

Observations of two hierarchical systems, consisting of an A-type
primary and a tight low-mass binary pair in a wide orbit allowed us
to extend this analysis to the lower mass regime. The primary of the
triple HIP 44127 suggests a young system age ($<$100 Myr), making the
M-dwarf pair interesting for comparison with evolutionary models. We have also shown that a
dynamical system mass significantly higher than the theoretical system
mass is suggestive of an unresolved spectroscopic component
within the system. Demonstrated on several known spectroscopic
binaries, systems which exhibit this discrepancy are ideal candidates
for future spectroscopic or interferometric observations in an attempt
to detect these hypothetical companions. Interferometric
  observations may be required to resolve these purported companions,
  as the rotationally broadened spectral lines of the rapidly rotating
  members of the monitoring sample \citep{Abt:1995eo} may preclude the
  detection of additional components using spectroscopy. Including
the three systems with evidence of an unresolved close companion, a
lower-limit on the higher-order multiplicity can be estimated from the
13 systems within the orbit subsample as 39 per cent double, 46 per
cent triple, and 15 per cent quadruple. The frequency of
  single A-type stars will be explored in an upcoming publication
  within the VAST survey paper series. The remaining systems for
which an orbit could not be determined are candidates for orbital monitoring
projects with ground-based high resolution observations. A number of
these systems are of particular interest, based on age estimates
derived from the position of the primary on the CMD, and the magnitude
difference between the two components.

\section*{Acknowledgements}
The authors wish to express their gratitude for the constructive
comments received from the referee, H. A. Abt. The authors also wish
to express their gratitude to A. Tokovinin, B. Mason, and W. Hartkopf
for comments which helped improve the paper. We gratefully acknowledge
several sources of funding. R. J. DR. is funded through a studentship
from the Science and Technology Facilities Council (STFCT) (ST/F
007124/1). J. P. is funded through support from the Leverhulme Trust
(F/00144/BJ) and the STFC (ST/F003277/1,
ST/H002707/1). R. J. DR. gratefully acknowledge financial support
received from the Royal Astronomical Society to fund collaborative
visits. Portions of this work were performed under the auspices of the
U.S. Department of Energy by Lawrence Livermore National Laboratory in
part under Contract W-7405-Eng-48 and in part under Contract
DE-AC52-07NA27344, and also supported in part by the NSF Science and
Technology CfAO, managed by the UC Santa Cruz under cooperative
agreement AST 98-76783. This work was supported, through J. R. G., in
part by University of California Lab Research Program
09-LR-118057-GRAJ and NSF grant AST-0909188. Based on observations
obtained at the Canada-France-Hawaii Telescope (CFHT) which is
operated by the National Research Council of Canada, the Institut
National des Sciences de l'Univers of the Centre National de la
Recherche Scientifique of France, and the University of Hawaii. Based
on observations obtained at the Gemini Observatory, which is operated
by the Association of Universities for Research in Astronomy, Inc.,
under a cooperative agreement  with the NSF on behalf of the Gemini
partnership: the National Science Foundation (United States), the
Science and Technology Facilities Council (United Kingdom), the
National Research Council (Canada), CONICYT (Chile), the Australian
Research Council (Australia),  Minist\'{e}rio da Ci\^{e}ncia e
Tecnologia (Brazil) and Ministerio de Ciencia, Tecnolog\'{i}a e
Innovaci\'{o}n Productiva (Argentina). The authors also wish to extend
their gratitude to the staff at the Palomar Observatory and the
UCO/Lick Observatory for their support and assistance provided during
the course of the observations. This research has made use of the
SIMBAD database, operated at CDS, Strasbourg, France. This publication
makes use of data products from the Two Micron All Sky Survey, which
is a joint project of the University of Massachusetts and the Infrared
Processing and Analysis Center/California Institute of Technology,
funded by the National Aeronautics and Space Administration and the
National Science Foundation. This research has made use of the
Washington Double Star Catalog maintained at the U.S. Naval
Observatory.
\bibliographystyle{mn2e}
\bibliography{Papers}
\label{lastpage}
\end{document}